\documentclass[%
 aip,
 jmp,%
 amsmath,amssymb,
 reprint,%
]{revtex4-2}

\usepackage{graphicx}
\usepackage{dcolumn}
\usepackage{bm}

\usepackage[colorlinks=true,hyperfootnotes=true,breaklinks=true,citecolor=blue]{hyperref}

\usepackage{algorithm2e}
\SetKwComment{Comment}{/* }{ */}

\DeclareMathOperator{\erf}{erf}

\begin{document}


\title[Gate--Level Statistical Timing Analysis: Exact Solutions, Approximations and Algorithms]{Gate--Level Statistical Timing Analysis:\\Exact Solutions, Approximations and Algorithms}
\thanks{This work has emanated from research supported in part by Synopsys, Ireland, and a research grant from Science Foundation Ireland (SFI) and is co-funded under the European Regional Development Fund under Grant Number 13/RC/2077.}

\author{Dmytro Mishagli}
\email{dmytro.mishagli@gmail.com}
\author{Eugene Koskin}%

\author{Elena Blokhina}%
 \email{elena.blokhina@ucd.ie}
\affiliation{ 
School of Electrical and Electronic Engineering, \\University College Dublin, Belfield, Dublin 4, \\Dublin, Ireland
}%


\date{\today}

\begin{abstract}
In this paper, the Statistical Static Timing Analysis (SSTA) is considered within the block--based approach. The statistical model of the logic gate delay propagation is systematically studied and the exact analytical solution is obtained, which is strongly non-Gaussian. The procedure of handling such (non-Gaussian) distributions is described and the corresponding algorithm for the critical path delay is outlined. Finally, the proposed approach is tested and compared with Monte Carlo simulations. 
\end{abstract}

\maketitle

\tableofcontents


\section{Introduction}
\label{sec:introduction}
Timing verification has never been more important for the design of digital integrated circuits (ICs) than now. Indeed, fabricated circuits must safely operate under working conditions and should satisfy the designed timing constraints. With the decrease of feature size, the gap between designs and fabricated circuits widens significantly. This is not only due to the increasing complexity of designs, but also because of variations of the parameters. At the moment, the technology has already scaled down below 5~nm getting close to physical limits. At such scales, the occurring variations of feature parameters (e.g. channel length) are of order of their nominal values. The traditional methods of the timing verification, static timing analysis (STA) and Monte Carlo (MC) simulations, no longer give satisfactory results. STA has become too pessimistic as it relies on utilising so-called corner values of the parameters, i.e. takes the worst case scenario ($3\sigma$ deviation from nominal values) even if the probability of it to happen is negligible in practice. At the same time, MC, while remains the most reliable tool, can take enormous time to finish a single design. All this moves research within statistical static timing analysis (SSTA) to front.

The main advantage of SSTA (and its curse at the same time) is that all design's parameters are considered as random variables (RVs) from the very beginning. It is especially important now as the impact of truly random variations (such as dopant concentration or oxide thickness) has become significant. Thus, the use of statistical approaches is natural. The main goal and task of SSTA is to determine a final distribution, probability density function (PDF) or cumulative density function (CDF), of a design slack, a so-called critical path delay.

From the analysis point of view, an IC is a so-called \emph{timing graph}~\cite{sapatnekar2004Timing}. Hence, the problem of timing analysis (either STA or SSTA) is equivalent to a graph optimisation problem, where one is required to find a shortest (longest) path in a graph~\cite{gerez1998AlgorithmsVLSIDesign}. There are two general approaches to the problem:
\begin{itemize}
  \item[(i)] block-based methods, which consider a delay propagation iteratively, level by level, from source to sink;
  \item[(ii)] path-based methods, within which a timing graph is considered as a set of paths.
\end{itemize}
For the reasons that will become clear further in this paper, the most of researchers' activity was carried out within the block-based approach.

\begin{figure*}[!t]
  \centering
    \includegraphics[width=0.95\textwidth]{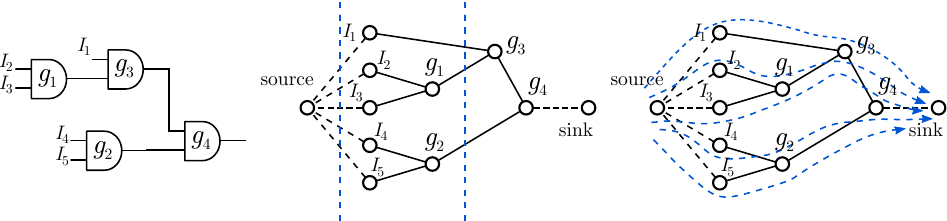}
  \caption{Combinational Logic Circuit modelling in Static Timing Analysis. A simple logic circuit on the left, and its timing graph in the middle and on the right. The timing graph can be traversed either within block-based or path-based approach.}
  \label{fig:dag}
\end{figure*}

The statistical domain brings the problem to another level of complexity, graph optimisation under uncertainties. This rises new challenges, such as computing a maximum of correlated non-Gaussian RVs. These challenges as well as various attempts to mitigate them have been described in two excellent reviews by Blaauw, Chopra \textit{et al}~\cite{blaauw2008StatisticalTimingAnalysis} and Forzan and Pandini~\cite{forzan2009StatisticalStaticTiming}, devoted to the first decade of SSTA (and mainly to block-based approaches). One of such challenges, which is of current interest, is the non-Gaussian nature of delays. Logic gate operation time is determined by the \emph{latest} of its arrival signals. In other words, nodal delay calculation requires computation $\max(\ldots)$ of all arrival signals' times. Even assuming the signals to have Gaussian delay distributions, one faces a problem that the result of maximum operation is a non-Gaussian PDF. Traditionally, at this stage simplifications and$/$or approximations are introduced that ineluctably bring errors to the delay computation. Moreover, it is widely believed in the community that exact solution does not exist (see e.g. \cite{lewis2019SpatialTimingAnalysis}).

This paper concludes our research started in \cite{freeley2018StatisticalSimulationsDelay,mishagli2020RBFApproximationNonGaussian}. For the Gaussian arrival delays, an exact expression of the distribution of a logic gate delay neglecting correlations was presented without derivation in~\cite{freeley2018StatisticalSimulationsDelay}. In principle, any non-Gaussian PDF can be represented as a mixture of Gaussian kernel functions, which allows one to apply the expression and propagate the delay with almost no loss in information. Such a representation requires solving of a corresponding optimisation problem. This was then investigated in~\cite{mishagli2020RBFApproximationNonGaussian}, where it was shown that a concept of a \emph{Gaussian comb} can give error in mean compare to MC less than $0.01\%$. In this paper, we present derivation of the logic gate delay distribution in a general case, taking correlations between nodes into account. We also investigate first four moments of the distribution: mean, standard deviation, skewness and kurtosis. Having full mathematical description of a system, we then discuss graph traversal algorithm that includes solution to an optimisation problem. Finally, we verify our approach by considering several application scenarios.

The paper is organised as follows. In Section~\ref{sec:background} we give details on the background and highlight the most common approaches and latest results in the field. The general statement of the problem is given in Section~\ref{sec:statement}. The core of the paper is in the next sections. Hence, we give the derivation of the gate delay distribution in Section~\ref{sec:distribution}. The algorithm for decomposition of non-Gaussian distributions and timing graph traversal is presented in Section~\ref{sec:gakeda}. Section~\ref{sec:discussion} gives discussion of the obtained results.

\section{Mathematical Preliminaries}
\label{sec:math}

In this work we will use the following explicit notation for the Gaussian PDF:
\begin{equation}\label{eq:gauss_pdf}
  \phi(x|\mu,\sigma) \stackrel{\text{def}}{=} \frac{1}{\sqrt{2\pi}\sigma} \varphi \left( \frac{x-\mu}{\sigma} \right), \quad \varphi(x) \stackrel{\text{def}}{=} e^{-\frac12 x^2},
\end{equation}
where the function $\varphi(x)$ is referred to as the \emph{Gaussian kernel}. The CDF of the \emph{standard} Gaussian distribution will be denoted as follows:
\begin{equation}\label{eq:gauss_cdf_1}
  \Phi(x) \stackrel{\text{def}}{=} \frac{1}{\sqrt{2\pi}} \int\limits_{-\infty}^x \varphi(x') dx' = \frac12 \left[ 1 + \erf \left( \frac{x}{\sqrt{2}} \right) \right],
\end{equation}
where the standard definition of the error function is used:
\begin{equation}\label{eq:error_function_definition}
  \erf(x) = \frac{2}{\sqrt{\pi}} \int\limits_0^x e^{-x'^2} dx'.
\end{equation}
For a compact notation, we will use
\begin{equation}\label{eq:gauss_cdf_2}
  F_i(x) \stackrel{\text{def}}{=} \Phi \left( \frac{x-\mu_i}{\sigma_i} \right) = \frac12 \left[ 1 + \erf \left( \frac{x-\mu_i}{\sqrt{2}\sigma_i} \right) \right].
\end{equation}

Let the function $g(z)$ be the PDF of the maximum $\zeta_2 = \max(X_1,X_2)$ for  two independent RVs $X_1$ and $X_2$ distributed normally. This PDF can be easily found. In this case, we note that the PDF and CDF can be factorised: $f(x_1, x_2) = f(x_1) f(x_2)$ and $F(x_1,x_2) = F(x_1)F(x_2)$. 
In this case, we have:
\begin{equation}\label{chap2:eq:max_2_independent}
  g(z) = f_1(z)F_2(z) + f_2(z)F_1(z),
\end{equation} 
where $f_i(x)$ and $F_i(x)$ are a Gaussian PDF and CDF of the $i^{\text{th}}$ distribution ($i=1,2$) respectively.

The PDF for the maximum of two \emph{correlated} Gaussian RVs has the form:
\begin{equation}\label{chap2:eq:max_2_correlated}
  \begin{aligned}
    g(z,\rho) &= f_1(z) \Phi\left[ \frac{1}{\sqrt{1-\rho^2}} \left( \frac{z-\mu_2}{\sigma_2} -\rho \frac{z-\mu_1}{\sigma_1} \right) \right]
    \\
    &+ f_2(z) \Phi\left[ \frac{1}{\sqrt{1-\rho^2}} \left( \frac{z-\mu_1}{\sigma_1} -\rho \frac{z-\mu_2}{\sigma_2} \right) \right].
  \end{aligned}
\end{equation}

\section{Background and State of the Art}
\label{sec:background}

In this section, we briefly outline the principles of SSTA and overview traditional techniques as well as recent papers.


In STA, whether it is deterministic or statistical, a combinational logic circuit is modelled with a timing graph, as it is shown in Fig.~\ref{fig:dag}. The graph is a Direct Acyclic Graph, i.e. it is given by a set $V$ of vertices and a set $E$ of edges, $G(V,E)$. Vertices correspond to nodes (logic gates) while edges describe interconnect delays. As we noted in Introduction, there are two methods to traverse the timing graph, path- and block-based. Within the path-based method, the total circuit delay equals to the maximal path delay, namely, for a circuit with $n$ paths, we have
\begin{equation}
  D = \max(\tau_1, \tau_2, \ldots, \tau_n),
\end{equation}
where $\tau_i$ is an accumulated delay in a path $i$. In SSTA, all $\tau_i$ are RVs and, in general case, are correlated and non-Gaussian. The distribution of the circuit delay $D$ in such a case is a non-trivial problem and remains unsolved. We will address this issue in a separate study.

Within the block-based method, the graph is traversed in a levelised manner, i.e. the delay is propagated from source to sink of a graph and computed for each block (see Fig.~\ref{fig:dag}). One cannot propagate further in a graph until all delays are computed within a block. The maximum transition time is determined by the latest arrival signal, thus, for a gate with two inputs, we have
\begin{equation}\label{eq:gate_delay_1}
  \ldots+\max(\tau_1,\tau_2) + \tau_{\text g} + \tau_{\text{int}} + \ldots,
\end{equation}
where $\tau_1$ and $\tau_2$ are the arrival times, $\tau_{\text g}$ is a gate's characteristic operation time, $\tau_{\text{int}}$ is an interconnect delay. All other delays like a wire delay $\tau_{\text{w}}$ can be added to \eqref{eq:gate_delay_1} and the sum can be continued. We will represent an output of a logic gate as follows:
\begin{equation}\label{eq:gate_delay_2}
  D_{\text{gate}} = \max(\tau_1,\tau_2) + \tau_0,
\end{equation}
where $\tau_0$ is a delay associated with interconnects, wires, etc. Thus, the block-based approach requires computation of \eqref{eq:gate_delay_2} at each node of a graph. In SSTA $\tau_i$ are RVs, and this expression is a source of all challenges of the approach.

Most of the works on SSTA are performed with the block--based approaches as attacking the SSTA problem~\eqref{chap3:eq:ssta_problem} is computationally hard because the total number of paths in IC grows exponentially. The research is focused on developing models of delays with further incorporating the models into algorithms for SSTA. One can distinguish three categories of the delay models:
\begin{enumerate}
    \item[(i)] \textit{Phenomenological}, or \textit{non-parameterised}, models. The approaches based on these models assume distributions of delays to be given (\textit{e.g.} from pre-characterisation of circuits elements using SPICE) and propose a framework to propagate the distributions through the timing graph $G$.
    \item[(ii)] \textit{Parameterised} models. These approaches model delays as functions of PVT variations and work directly with RVs. Information regarding the variations can be obtain from foundries or \textit{via} MC simulations. As their final step, the methods need to relate the models to particular distributions. Techniques like AWE~\cite{pillage1990AsymptoticWaveformEvaluation,chiprout1994AsymptoticWaveformEvaluation} can be used for this.
    \item[(iii)] \textit{Microscopic} models aim to establish relations between the PVT variations and delays in transistors and, hence, logic gates. These models are developed for better understanding of the problem and for more accurate parameterised models
\end{enumerate}

Below we review these models in detail.

\subsubsection{Phenomenological delay models}

A pioneering work on SSTA was done by Berkelaar~\cite{berkelaar1997StatisticalDelayCalculation}, where it was proposed to consider distributions of the arrival times' delays of a logic gate, \textit{i.e.} $\tau_1$ and $\tau_2$ in \eqref{chap3:eq:gate_delay_2}, as independent Gaussian RVs. Thus, the distribution of $\max(\tau_1,\tau_2)$ is given by \eqref{chap2:eq:max_2_independent}. It was proposed then to approximate the mean value of a gate delay simply by adding the mean value of $\max(\tau_1,\tau_2)$ to the mean value of $\tau_0$, thus, treating the distribution \eqref{chap2:eq:max_2_independent} as if it was a Gaussian one. The standard deviation of a gate delay is calculated as $\sigma^2_{\text{gate}} = \sigma^2_{\max}+\sigma^2_0$. After these calculations performed, the gate delay was considered as a Gaussian RV, \textit{i.e.} its distribution given by \eqref{eq:gauss_pdf}, and the whole process was repeated at the next gate. In principle, this repeats the famous Clark's algorithm~\cite{clark1961GreatestFiniteSet}, but for a case of zero correlations. The exact expression for \eqref{chap2:eq:max_2_independent}, as well as for the mean value and standard deviation of an RV given by this distribution, were presented later, in a joint work with Jacobs in \cite{jacobs2000GateSizingUsing}, where this delay model was applied to a gate sizing problem.

In these works, \cite{berkelaar1997StatisticalDelayCalculation} and \cite{jacobs2000GateSizingUsing}, correlations between delays were not taken into account. Tsukiyama~\textit{et al.} addressed this issue in~\cite{tsukiyama2000NewStatisticalStatic,tsukiyama2001StatisticalStaticTiming}. It was proposed to use the known expressions for the PDF of two correlated Gaussian RVs, namely, equation~\eqref{chap2:eq:max_2_correlated}. Other steps were the same as in works by Berkelaar and Jacobs. Thus, for each gate (a node in a timing graph), the exact distribution~\eqref{chap2:eq:max_2_correlated} was approximated with a Gaussian distribution by matching two first moments and taking into account the correlation coefficient, as per Clark~\cite{clark1961GreatestFiniteSet}. Such an algorithm has a complexity of $\mathcal{O}(|V|\cdot|E|)$, as it was reported in~\cite{tsukiyama2001StatisticalStaticTiming}. Hence, the approach by Berkelaar~\cite{berkelaar1997StatisticalDelayCalculation} has $\mathcal{O}(|V|+|E|)$ time complexity, as the delays are independent.

Azuma~\textit{et al.}~\cite{azuma2017ApproximatingMaximumGaussians} have proposed to approximate the PDF of the maximum of two Gaussian distributions \textit{via} GMM. The authors gave closed formulas for the parameters of the GMM. It should be noted that the proposed approach required that the distributions one takes the maximum of are either Gaussians or given in the GMM form. However, it needs to be explained how an arbitrary non-Gaussian distribution should be decomposed into the mixture. Therefore, assuming the initial delays are distributed normally, the overall complexity with respect to the total number of atomic operations, $N$, is $\mathcal O(m^2N)$, where $m$ is the number of components in the GMM.

Chen~\textit{et al.}~\cite{chen2015InverseGaussianDistribution} proposed to approximate non-Gaussian distributions by an inverse Gaussian distribution, which has a shape parameter responsible for skewness. Such an approximation is meant to be done as a preprocessing step, while the actual distributions can be retrieved from prior simulations (this can be done only ones). Then, the traversal of these non-Gaussian distributions can be done (hence, convolution) \textit{via} summation of the parameters of the inverse Gaussian PDFs. The proposed approach works well when the $\max$ is not required, \textit{e.g.} for a chain of inverters of for a cascade of AND (NAND) gates, when both inputs are equal (thus, $\max$ is trivial). This makes the approach not suitable for real VLSI designs.

A very recent work by Jin~\textit{et al.}~\cite{jin2022StatisticalCellDelay} uses a so-called log--extended--skew--normal distribution to represent gate delays. This distribution has four parameters to match not only the mean and standard deviation, but also the skewness and kurtosis. Though it has remarkable agreement with MC simulations, it has the same limitations as the work by Chen~\textit{et al.}~\cite{chen2015InverseGaussianDistribution}, \textit{i.e.} it is unclear how the $\max$ operation should be performed.

\subsubsection{Parameterised delay models}

Seminal works within the block--based approach were made by Chang \& Sapatnekar~\cite{chang2003StatisticalTimingAnalysis,chang2005StatisticalTimingAnalysis}, Chang~\textit{et al.}~\cite{chang2005ParameterizedBlockbasedStatistical}, and Visweswariah~\textit{et al.}~\cite{visweswariah2004FirstOrderIncrementalBlockBased,visweswariah2006FirstOrderIncrementalBlockBased}. These works continued the approach of Berkelaar to use the moment matching Clark's method~\cite{clark1961GreatestFiniteSet} and have shaped block--based SSTA. The works by Visweswariah~\textit{et al.}~\cite{visweswariah2004FirstOrderIncrementalBlockBased,visweswariah2006FirstOrderIncrementalBlockBased} resulted in an IBM's tool \texttt{EinsStat} for SSTA. Let us consider these papers in detail.

It was proposed to consider a gate delay as a function of process variations. Thus, for $n$ sources of variations $X_i$, the gate delay can be written in a general form as
\begin{subequations}\label{chap3:eq:canonical_form}
\begin{equation}\label{chap3:eq:canonical_form_a}
  d = d(X_1,X_2,\ldots,X_n),
\end{equation}
where the exact form of the r.h.s. depends on a delay model used. Thus, Chang \& Sapatnekar~\cite{chang2003StatisticalTimingAnalysis,chang2005StatisticalTimingAnalysis} proposed a direct extension of \cite{berkelaar1997StatisticalDelayCalculation,jacobs2000GateSizingUsing,tsukiyama2000NewStatisticalStatic,tsukiyama2001StatisticalStaticTiming} by transforming a set of correlated RVs into a set of i.i.d. that are distributed normally, \textit{i.e.} $\sim \mathcal N(0,1)$, by PCA. In this case, \eqref{chap3:eq:canonical_form_a} is a linear combination of variation parameters. For such linear combinations, the Clark's algorithm~\cite{clark1961GreatestFiniteSet} is then used.

A notable result of \cite{chang2003StatisticalTimingAnalysis,chang2005StatisticalTimingAnalysis} is a method to extract correlations. Chang \& Sapatnekar proposed a grid model for spatial correlations, where a die is to be split into $n$ grids, and the correlation between circuit elements that are in the grids decays with the distance between the grids. The overall complexity of the algorithm is $\mathcal O(pn(|E|+|V|))$, where $p$ is a number of variation parameters considered and $n$ is a number of grids. It also should be noted that a quad--tree method for modelling the spatial correlations on a die was proposed by Agarwal~\textit{et al.}~\cite{agarwal2003StatisticalDelayComputation}. As the name suggests, the die is divided into levels, and for each level $i$, the die is partitioned into $2^{(i)}$-by-$2^{(i)}$ squares, where the square for the first level has an area of the whole die, and the last level is split into 16 squares. An independent RV is then associated with each square. These grid and quad--tree methods are often employed in different parameterised block--based approaches. Further information can be found in a review by Forzan \& Pandini~\cite{forzan2009StatisticalStaticTiming}.

At the same time, Visweswariah~\textit{et al.}~\cite{visweswariah2004FirstOrderIncrementalBlockBased,visweswariah2006FirstOrderIncrementalBlockBased} proposed a linear \emph{canonical model} of a delay. Thus, the general delay~\eqref{chap3:eq:canonical_form_a} was written as follows:
\begin{equation}\label{chap3:eq:canonical_form_b}
  d = a_0 + \sum\limits_{i=1}^n a_i \Delta X_i + a_{n+1} \Delta R,
\end{equation}
where $a_0$ is the mean or nominal value, $\Delta X_i$ represent the variation of $n$ global sources of variation $X_i$ from their nominal values, $a_i$ are sensitivities to each of the RVs, and $\Delta R$ is the variation of an independent RV $R$. The sensitivities can be obtained, for example, from circuit simulations when a chip is characterised. The Clark's method~\cite{clark1961GreatestFiniteSet} was written in terms of probabilities, using a concept of \emph{tightness probability}, $T_A=P(A>B)$, a probability that arrival time $A$ is greater than $B$. Thus, closed--form expressions for a linear approximation of two canonical forms~\eqref{chap3:eq:canonical_form_b} was obtained. At this point, these works are similar to~\cite{chang2003StatisticalTimingAnalysis,chang2005StatisticalTimingAnalysis}, with the only difference that the variance of the approximation of two linear forms is computed differently. This approach has linear complexity with a circuit size, $\mathcal O(n \cdot N)$, where $N$ is the total number of $\max$ and $\mathrm{sum}$ operations.

The linear canonical model~\eqref{chap3:eq:canonical_form_b} used in the discussed above approaches, is one of the main sources of inaccuracies, which was quickly understood. This is because the linearity of \eqref{chap3:eq:canonical_form_b} implies Gaussian distribution for delays and the $\max$ operation. Therefore, many attempted to address this problem, remaining in a paradigm of the canonical forms~\eqref{chap3:eq:canonical_form_a}. Thus, Chang~\textit{et al.}~\cite{chang2005ParameterizedBlockbasedStatistical} proposed the \emph{generalised canonical form}
\begin{equation}\label{chap3:eq:canonical_form_c}
  d = a_0 + \sum\limits_{i=1}^n a_i \Delta X_i + a_{n+1} \Delta R + f_d(\Delta X_N),
\end{equation}
where $f_d(\Delta X_N)$ represents dependencies of $d$ on nonlinear (hence, non-Gaussian) parameters. Chang~\textit{et al.} proposed to use numerically computed tables to describe $f_d(\Delta X_N)$.

Alternative to adding non-linear terms to the canonical form approach was proposed by Singh and Sapatnekar~\cite{singh2006StatisticalTimingAnalysis,singh2008ScalableStatisticalStatic}. The delay was still represented using the linear canonical model~\eqref{chap3:eq:canonical_form_a} where non-Gaussian RVs were given by another term. The non-Gaussian term is then addressed by \emph{preprocessing} the random vector of correlated random variables \textit{via}~\cite{manduchi1999IndependentComponentAnalysis,hyvarinen2000IndependentComponentAnalysis,murphy2022ProbabilisticMachineLearning}. The correlated Gaussian part of~\eqref{chap3:eq:canonical_form_a} is handled with PCA as in~\cite{chang2003StatisticalTimingAnalysis,chang2005StatisticalTimingAnalysis}. After the preprocessing step, which is supposed to be made only once, hence, it does not contribute to the total complexity of the algorithm, the moments of the transformed RVs are calculated. The PDFs and CDFs are restored from the canonical forms using numerical techniques proposed in~\cite{li2004AsymptoticProbabilityExtraction}; the techniques uses \textit{Pad\'e Approximants} and requires computation of moments of the canonical forms.

In contrast to \cite{chang2005ParameterizedBlockbasedStatistical}, Zhan~\textit{et al.}~\cite{zhan2005CorrelationawareStatisticalTiming} and Zhang~\textit{et al.}~\cite{zhang2005CorrelationpreservedNonGaussianStatistical} introduced quadratic terms into \eqref{chap3:eq:canonical_form_b} instead of a function $f_D(\Delta X_N)$. However, such an approximation of the $\max$ still introduces noticeable errors. For $N$ being a circuit size, the worst--case timing complexity of both approaches is $\mathcal O(n^2 N)$, where $n$ is a number of terms in the quadratic canonical form. The authors indicated that the complexity can be reduced to $\mathcal O(n \cdot N)$ if the most of the cross terms ignored. At the same time, \cite{zhan2005CorrelationawareStatisticalTiming} uses expensive numerical integration for computation of the $max$ of the canonical forms, and \cite{zhang2005CorrelationpreservedNonGaussianStatistical} requires additional simplifications. Therefore, these approaches do not give advantage compare to the linear canonical model.

\begin{table}[!t]
\begin{center}
\begin{tabular}{ lp{25mm}lp{60mm} } 
    \hline
    \hline
    Reference & Complexity & Features & Comments \\
    \hline
    \cite{berkelaar1997StatisticalDelayCalculation,jacobs2000GateSizingUsing} & $\mathcal{O}(|V|+|E|)\phantom{\displaystyle\frac12}$ & G., $\rho=0$ & Eq.~\eqref{chap2:eq:max_2_independent} is approximated with a Gaussian by two moments matching  \\ 
    \cite{tsukiyama2000NewStatisticalStatic,tsukiyama2001StatisticalStaticTiming} & $\mathcal{O}(|V|\cdot|E|)$ & G., $\rho\neq0$ & Correlations are taken into account by utilising Eq.~\eqref{chap2:eq:max_2_correlated}; the result is approximated with a Gaussian by matching two moments. \\
    \cite{chang2003StatisticalTimingAnalysis,chang2005StatisticalTimingAnalysis} & $\mathcal{O}(pn \cdot N)$ & G., $\rho\neq0$ & Correlations are treated with PCA; $\max$ is approximated with a Gaussian by Clark's method~\cite{clark1961GreatestFiniteSet}; \newline $p$ is a number of variation parameters, $n$ is a number of grids.
    \\ \cite{chang2005ParameterizedBlockbasedStatistical,visweswariah2004FirstOrderIncrementalBlockBased,visweswariah2006FirstOrderIncrementalBlockBased} & $\mathcal{O}(n \cdot N)$ & G., $\rho\neq0$ & Linear canonical form is used and, hence, $\max$ is approximated with a Gaussian using tightness probability; correlations are due to global sources of variations, $n$.
    \\
    \cite{singh2006StatisticalTimingAnalysis,singh2008ScalableStatisticalStatic} & $\mathcal{O}(n \cdot |E|)$ & non-G., $\rho\neq0$ & Linear canonical model is used, but non-Gaussian RVs are treated with ICA. Correlated RVs are transformed with PCA as in \cite{chang2003StatisticalTimingAnalysis,chang2005StatisticalTimingAnalysis}; pdfs are restored from the canonical forms using Pad\'e-Approximants-based methods~\cite{li2004AsymptoticProbabilityExtraction}.
    \\
    \cite{zhan2005CorrelationawareStatisticalTiming} & $\mathcal{O}(n^2N)$ \newline \rightline{$\approx \mathcal{O}(n \cdot N)$} & non-G., $\rho\neq0$ & Quadratic canonical form is used; $\max$ is computed using numerical techniques; Correlations are taken effectively, as in \cite{chang2005ParameterizedBlockbasedStatistical,visweswariah2004FirstOrderIncrementalBlockBased,visweswariah2006FirstOrderIncrementalBlockBased}; \newline $n$ is the same as above.
    \\  
    \cite{zhang2005CorrelationpreservedNonGaussianStatistical} & $\mathcal{O}(n^2N)$ \newline \rightline{$\approx \mathcal{O}(n \cdot N)$} & non-G., $\rho\neq0$ & Quadratic canonical form is used; Correlations and $\max$ are treated in the same manner as in \cite{chang2003StatisticalTimingAnalysis,chang2005StatisticalTimingAnalysis,chang2005ParameterizedBlockbasedStatistical,visweswariah2004FirstOrderIncrementalBlockBased,visweswariah2006FirstOrderIncrementalBlockBased}; \newline $n$ is the same as above.
    \\
    \cite{khandelwal2005GeneralFrameworkAccurate} & $\mathcal{O}(n^2N)$ & non-G., $\rho\neq0$ & Delays are represented \textit{via} Taylor series; $\max$ is represented via polynomials by linear regression; further matching of the three moments gives a new canonical form; \newline $n$ is the highest order of the polynomials. \\
    \hline
    \hline
\end{tabular}
\end{center}
  \caption[Complexity of block--based SSTA approaches from literature]{Complexity of block--based SSTA approaches from the literature. Here $|E|$ is a number of edges, and $|V|$ is a number of vertices in a timing graph; $N$ is a total number of $\max$ and $\text{sum}$ operations. Further details are in the text.}
  \label{chap3:tab:compexity_a}
\end{table}

\begin{table}[!t]
\begin{center}
\begin{tabular}{ lp{25mm}lp{60mm} } 
    \hline
    \hline
    Reference & Complexity & Features & Comments \\
    \hline
    \cite{cheng2007NonLinearStatisticalStatic,cheng2012FourierSeriesApproximation} & $\mathcal{O}(nK^3N)\phantom{\displaystyle\frac12}$ \newline \rightline{$\approx \mathcal{O}(nK^2 \cdot N)$} & non-G., $\rho\neq0$ & PDF of $\max$ of two quadratic canonical forms is approximated by \emph{Fourier series}; further matching of the three moments gives a new canonical form; \newline $K$ is the highest order of Fourier series, $n$ is a number of variation sources.
    \\
    \cite{cheng2008NonGaussianStatisticalTiming,cheng2009NonGaussianStatisticalTiming} & $\mathcal{O}(n^3N)$ \newline \rightline{$\approx \mathcal{O}(n \cdot N)$} & non-G., $\rho\neq0$ & Extension of \cite{cheng2007NonLinearStatisticalStatic,cheng2012FourierSeriesApproximation}; here PDF of $\max$ of two quadratic canonical forms is approximated by \emph{least--squares--error minimisation}; \newline $n$ is as above.
    \\
    \cite{vijaykumar2014StatisticalStaticTiming,ramprasath2016SkewNormalCanonicalModel} & $\mathcal O(n^2N)$ & non-G., $\rho\neq0$ & PDF of $\max$ and resulting delays are modelled with skew--normal distribution; delays are propagated \textit{via} moment matching;
    \newline $n$ is as above.
    \\
    \cite{shebaita2018GeneralizedMomentBased} & $\mathcal{O}(m^2 N)$ & non-G., $\rho=0$ & Delays are represented \textit{via} moments, and atomic operations are applied to them; PDF is restored via AWE method; \newline $m$ is the number of moments used.
    \\
    \cite{chen2015InverseGaussianDistribution} & NA & non-G., $\rho=0$ & Non-G. is approximated by an inverse Gaussian distribution, and the convolution is done via summation of the parameters of such distributions; \newline $\max$ \emph{is not addressed}.
    \\
    \cite{azuma2017ApproximatingMaximumGaussians} & $\mathcal O(m^2N)$ & G., $\rho\neq0$ & $\max$ of two Gaussians is approximated with a GMM with further model reduction; \newline $m$ is the number of components in GMM.
    \\
    \cite{lewis2019SpatialTimingAnalysis} & $\mathcal{O}(nm \cdot N)$ & G., $\rho\neq0$ & Linear canonical delay model is used; $\max$ is performed as pairwise comparison of polynomials; \newline $n$, $m$ are a total numbers of polynomials in a fanin set for a node and in its delay set respectively. \\
    \hline
    \hline
\end{tabular}
\end{center}
  \caption[Complexity of block--based SSTA approaches from literature (cont.)]{Complexity of block--based SSTA approaches from the literature (continuation of Table~\ref{chap3:tab:compexity_a}). Here $N$ is a total number of $\max$ and $\text{sum}$ operations. With ``NA'' marked those approaches where the total complexity is unknown since the authors did not propose the full SSTA algorithm. Further details are in the text.}
  \label{chap3:tab:compexity_b}
\end{table}

A different approach was proposed by Khandelwal and Srivastava~\cite{khandelwal2005GeneralFrameworkAccurate}, where the authors did not rely on the canonical form~\eqref{chap3:eq:canonical_form} explicitly. Instead, gate delays and arrival times were presented as polynomials using a Taylor series expansion. With the second order degree, this approach is similar to \eqref{chap3:eq:canonical_form_c} though. The maximum of two polynomials was approximated with a polynomial as well by linear regression, instead of the moment matching technique. However, the regression requires a MC simulation in the inner loop of the proposed procedure. The spatial correlations are encountered effectively as coefficients of the polynomials with the grid--based correlation model employed. Thus, the proposed method scales linearly with the total number of $\max$ and $\mathrm{sum}$ operations, and it is quadratic with respect to the degree of polynomials used, $\mathcal O(n^2N)$. But the hidden MC simulation makes this approach unlikely to be used for real VLSI designs.

The above discussed approaches either were inaccurate (since rely on a linear model of a delay) or did not scale well due to usage of expensive numerical methods. Therefore, Cheng~\textit{et al.} attempted to mitigate these disadvantages in a series of works, \cite{cheng2007NonLinearStatisticalStatic,cheng2012FourierSeriesApproximation} and \cite{cheng2008NonGaussianStatisticalTiming,cheng2009NonGaussianStatisticalTiming}. The first approach~\cite{cheng2007NonLinearStatisticalStatic,cheng2012FourierSeriesApproximation} proposes another quadratic canonical model of delay:
\begin{equation}\label{chap3:eq:canonical_form_d}
    D = d_0 + \sum_i^n  (a_i X_i + b_iX_i^2) + a_{\text r} X_{\text r} + b_{\text r} X_{\text r}^2,
\end{equation}
\end{subequations}
where $d_0$ is nominal delay (mean value), $X_i$ is an RV that represents global sources of variation and $X_{\text r}$ is a purely independent random variation. The RVs $X_i$ allowed to follow arbitrary distribution, and the correlations are proposed to be addressed \textit{via} ICA. The bottleneck of the method is the PDF of the $\max$ operation applied to the canonical forms \eqref{chap3:eq:canonical_form_d}, which is computed using Fourier series. Thus, the overall complexity of this method with respect to a total number of atomic operations, $N$, is $\mathcal{O}(nK^3N)$, where $n$ is a total number of variation sources and $K$ is the highest order of Fourier series. Then, the complexity was decreased to $\mathcal O(nK^2N)$ in \cite{cheng2009NonGaussianStatisticalTiming}.

The second study of Cheng~\textit{et al.}~\cite{cheng2008NonGaussianStatisticalTiming,cheng2009NonGaussianStatisticalTiming} goes further towards reducing the complexity of \cite{cheng2007NonLinearStatisticalStatic}. It was proposed a different and more efficient way of constructing PDFs from the maximum of two canonical forms. Similar to \cite{zhang2006NonGaussianStatisticalParameter}, the non-Gaussian RVs were modelled with a quadratic polynomials. The $\max$ was approximated with another quadratic polynomial, $h(A,B,\Theta)$, where $A$, $B$ are non-Gaussian RVs written in a quadratic canonical form and $\Theta$ is a set of three fitting parameters. Then, these parameters were found from an optimisation problem of minimising a quadratic error between $h(A,B,\Theta)$ and $\max(A-B,0)$, while matching the mean value of the $\max$ operation. There was proposed a closed--from solution to the optimisation problem, thus, this gave a significant speed up compare to~\cite{cheng2007NonLinearStatisticalStatic,cheng2012FourierSeriesApproximation}. For the total number of $n$ variational sources, the overall complexity of this method is $\mathcal O(n^3N)$, however, it can be reduced to $\mathcal O(n \cdot N)$ if the cross--terms in the canonical forms are ignored similar to~\cite{zhan2005CorrelationawareStatisticalTiming,zhang2005CorrelationpreservedNonGaussianStatistical}.

A different approach was initiated by Chopra~\textit{et al.}~\cite{chopra2006NewStatisticalMax}, who proposed to take into account skewness of the $\max$ operation by modelling it with a skew--normal distribution~\cite{azzalini1985ClassDistributionsWhich,azzalini2013SkewNormalRelatedFamilies}, which PDF is $f(x;\lambda)=2\phi(x)\Phi(\lambda x)$, where $\lambda$ is a shape parameter that determines the skewness. The authors approximated the $\max$ of two linear canonical forms~\eqref{chap3:eq:canonical_form_b} with the skew--normal distribution by moment--matching techniques, similar to Clark's~\cite{clark1961GreatestFiniteSet}, and derived corresponding formulas to take into account the third moment.

This direction was continued in the works by Vijaykumar \& Vasudevan~\cite{vijaykumar2014StatisticalStaticTiming} and Ramprasath \textit{et al.}~\cite{ramprasath2016SkewNormalCanonicalModel}, who extended this idea beyond the linear canonical form. Thus, the authors of \cite{ramprasath2016SkewNormalCanonicalModel} introduced the so-called skew-canonical representations of the quadratic canonical form~\eqref{chap3:eq:canonical_form_d} and its linear relaxation. The moments of the skew--normal approximation of the $\max$ are expressed \textit{via} Owen's $T$-function~\cite{owen1956TablesComputingBivariate}. The correlations were represented using the quad--tree model~\cite{agarwal2005StatisticalTimingBased}. The authors reported that the accuracy of their approach is comparable to that of Cheng's~\textit{et al.}~\cite{cheng2008NonGaussianStatisticalTiming,cheng2009NonGaussianStatisticalTiming} but the run time on average was two orders of magnitude lower than for quadratic models. This is due to the fact that this approach does not require the restoration of a PDF at each step neither by Fourier series nor by least--squares fitting. One can conclude that the complexity of the approach is comparable to quadratic models with moment matching approaches, $\mathcal O(n^2N)$.

Shebaita~\textit{et al.}~\cite{shebaita2018GeneralizedMomentBased} proposed to represent delays \textit{via} moments in terms of distributions' poles and residues. The authors gave general formulas for the moments given an arbitrary distribution. Obtaining the poles and residues of the distributions is the required pre-prepossessing step for the method. The resulting distributions, PDFs and CDFs, are proposed to be restored using the AWE method~\cite{pillage1990AsymptoticWaveformEvaluation,chiprout1994AsymptoticWaveformEvaluation}. However, the problem of correlations is not addressed in the paper, which limits the applicability of the approach. Given the number of moments used to represent delays, $m$, the overall timing complexity with respect to the total number of atomic operations, $N$, is $\mathcal O(m^2 \cdot N)$.

A recent work by Lewis \& Schmit \cite{lewis2019SpatialTimingAnalysis} aims to decrease the number of RVs used in parameterised block--based approaches. Using the linear canonical model~\eqref{chap3:eq:canonical_form_a} as a starting point, the delays are represented in terms of polynomial basis functions, which are then propagated through a timing graph. The information about spatial correlations is embedded into the polynomials from grid--based models of the correlations. The $\max$ is performed as pairwise comparison of the polynomials, and the summation results in merging the polynomials in a new one. However, the proposed methodology is limited only to Gaussian RVs, and the authors did not discuss possible extension to non-Gaussian distributions. The overall complexity with respect the total number of atomic operations, $N$, is $\mathcal O (nm \cdot N)$, where $n$ is a total number of polynomials in the fanin set for a node, and $m$ is a total number of polynomials in the delay set for a node.

Another recent work by Jin~\textit{et al.}~\cite{jin2019PreciseBlockBasedStatistical} aims to increase the accuracy of block--based approaches by treating the $\max$ operation using multivariate adaptive regression splines (MARS). The idea is to use SPICE simulations to obtain training data to build the MARS models. The result of the $\max$ operation is mapped onto a skew--normal distribution. The complexity of the algorithm with respect to the number of the atomic operations, $N$, is $\mathcal O (MN)$, where $M$ is the number of samples in the training data. Overall, while this approach has high accuracy, it requires exhaustive numerical computations at each step.

To summarise, we have seen that all the proposed methods scale linearly with the total number of the atomic ($\max$ and $\mathrm{sum}$) operations in a circuit, $N$, but with a scale factor, \textit{i.e.} $\mathcal O(k\cdot N)$ (see Tables~\ref{chap3:tab:compexity_a} and \ref{chap3:tab:compexity_b}). This scale factor, $k$, varies with approaches and can imply exhaustive numerical computations at each gate. Linearity with $N$ should not be surprising as all the methods use BFS algorithm, which complexity is $\mathcal O(|V| + |E|)\approx \mathcal O(N)$.

\subsubsection{Microscopic delay models}

There is a brunch of research dedicated to studying the sources of variations, which is of particular importance for effective delay models. The main focus is on developing the current--based models of delays of the logic cells~\cite{zolotov2007CompactModelingVariational}. Since these works aim to determine and$/$or model the fundamental stochastic behaviour of the process variations, we group them as \emph{microscopic} models, even though they do not consider explicitly the microscopic properties of matter such as distributions of particles. Developing a microscopic model was not aimed to be a part of the Thesis, thus, we will only briefly discuss some of the works.

Thus, Fatemi~\textit{et al.}~\cite{fatemi2006StatisticalLogicCell} modelled the cell output voltage waveform as a Markovian stochastic process considering the variations of physical parameters (such as channel width, length, and temperature) Gaussian. This allowed the authors to obtain a distribution of the cell delay, which was in better agreement with the MC simulations done in SPICE than the most recent current--based model at that time by Keller~\textit{et al.}~\cite{keller2004RobustCelllevelCrosstalk}. 

In parallel, Y.~Cao \& L.T.~Clark~\cite{cao2005MappingStatisticalProcess} presented a physically based, analytical delay variability model for both regions, at saturation and at sub-threshold. Starting from the well-known $\alpha$-power law for a gate delay in a MOSFET~\cite{sakurai1990AlphapowerLawMOSFET}, the authors took short--channel effects into account based on the previous work by Orshansky~\textit{et al.}~\cite{orshansky1999DirectSamplingMethodology}. That work allowed the authors to (i) model the dependence of $V_{\text{th}}$ on channel length, (ii) utilise a formula for the saturation of carrier velocity, and (iii) formulate a unified formula for the drive current.

Shinkai~\textit{et al.} \cite{shinkai2006GateDelayModel,shinkai2013GatedelayModelFocusing} considered a wide range of PVT variations and their influence on the gate delay. With the focus on fluctuations of the output current, $I_{\text d}$ the authors proposed a methodology to relate PVT variations to the gate delay and output slew given input slew and output load. It should be noted that the methodology is agnostic with respect to the gate delay model, thus, any of the models discussed in the previous subsection can be applied.

Sinha~\textit{et al.}~\cite{sinha2016PracticalStatisticalStatic} studied the possibility of practical usage of the current--source models in industrial timing analysis environments~\cite{zolotov2012TimingAnalysisNonseparable}. The authors have presented a technique for efficient storing the waveforms in memory and validated it for 14~nm technology designs.

P.~Cao~\textit{et al.}\cite{cao2019AnalyticalGateDelay} related analytical results from \cite{harris2010TransregionalModelNearthreshold,keller2014CompactTransregionalModel} for the current variations at sub- and near-threshold regions to the parameters of the log--skew--normal (LSN) distribution by moment matching. In other words, they gave microscopic reasoning for gate delays modelled with LSN distributions.

\section[Statement of the Problem]{Statement of the Problem in Block\,--\,Based SSTA}
\sectionmark{Statement of the Problem}
\label{sec:statement}

\begin{figure}[t!]
\centering
\includegraphics[width=\textwidth]{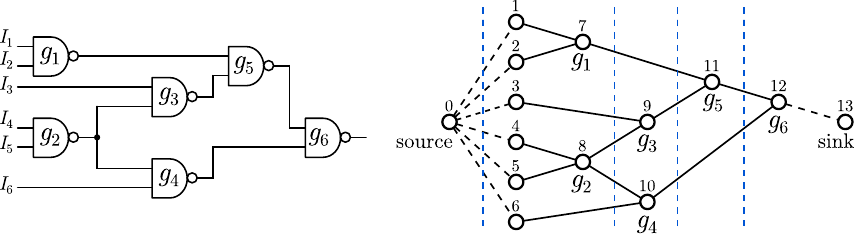}
\vspace{0.5cm}
\caption[Example of combinational logic circuit and its timing graph for block--based analysis]{An example combinational circuit with a symbolic notation of logic gates and interconnects. The timing graph is split in a levelised manner (blocks), which indicates the block--based analysis.}
\vspace{0.5cm}
\label{fig:circuit_graph}
\end{figure}

Consider a simple combinational logic circuit and its timing graph, as shown in Figure~\ref{fig:circuit_graph}. As was noted above, the edges of this graph represent logic gates, and the vertices represent the inputs and outputs of the logic gates.  Firstly, we note that logic gates have an `internal structure' since they are made of  transistors. This results in a finite characteristic time needed for gates to operate. This is one of the sources of delays in a circuit. Secondly, due to operational delays existing in every gate (and also in every  interconnect), signals, propagating in a logic circuit, arrive at the next logic gate also with delays. Most logic gates have at least two inputs. They need both inputs to `arrive' (\textit{i.e.} to change from Low to High or vice versa) before a logic operation can be performed on them and the output can be generated.  For this reason, the maximum of two (or more) input delays is another contribution to circuit delay. We will discuss both contributions and their mathematical form below. 

Thus, the main problem of block-based timing analysis for logic circuits can be formulated as the mathematical problem of calculating the $\max$ function of arrival times and accommodation the operation time of logic gates and interconnect in the overall delay calculation. In other words, this is a problem for graph optimisation. Graphs are typically described by corresponding adjacency matrices. Firstly, the 
BFS algorithm (see Figure~\ref{chap2:fig:pseudo-code_breadth}) is used to understand in what order a timing graph is traversed. As soon as the sequence of operations required to find the delay between two nodes is found, one must use gate delays and interconnect delays involved in this sequence to calculate the actual delay. 

Both contributions, the gate operation time and the  $\max$ function, can have a significant impact on the overall circuit delay. It is difficult to say which contribution is more important, and they both should be taken into account. In principle, it is quite straightforward to do in the context of deterministic timing analysis, but it is not the case when uncertainty arises. When it is necessary to include variations of parameters, SSTA is required. In such a case, the  arrival and gate operation times are described by random variables given by some distributions.

Looking at the mathematical statement of the problem, we recall that at the individual gate level, delay propagation is described by two \emph{atomic} operations: computing the maximum ($\max$) of two delays entering a gate and the summation ($\mathrm{sum}$) of the latter with the delay of the gate due to its operation time, the interconnect delay, the wire delay, etc. From the statistical point of view, when these operations are applied to RVs, the delay of a gate with two inputs reads
\begin{equation}\label{eq:gate_operation}
    \eta= \max(X_1,X_2) + X_0,
\end{equation}
where $X_1$ and $X_2$ are the RVs that describe the arrival times of input signals, and $X_0$ is the RV that gives the gate operation time.

In terms of distributions, combinations like \eqref{eq:gate_operation} are equivalent to the \emph{convolution} of two functions: the PDF $f_{\text{max}} (x)$ of the maximum  $\max(X_1, X_2)$ of the arrival signals' delays with the PDF $f_0(x)$ that gives the distribution of the gate operation time  $X_0$:
\begin{equation}\label{eq:convol}
  f_{\text{gate}} (x) \stackrel{\text{def}}{=} (f_{\text{max}} * f_0)(x) = \int\limits_{-\infty}^{\infty}f_{\text{max}}(x')f_0(x-x')dx'.
\end{equation}
This convolution formula is general and valid for any PDFs, which do not have to be Gaussian. On the other hand, this is the bottleneck of the gate-level analysis: expression~\eqref{eq:convol} can be calculated numerically (requiring computational resources and time). Thus, if this formula is applied numerically to every gate in a VLSI circuit, the analysis would not be completed in acceptable time. For this reason, the challenge here is to find an approach to approximate this formula and get some sort of an analytical expression for convolution~\eqref{eq:convol}.

\begin{figure}[t!]
\center
\includegraphics[width=0.80\textwidth]{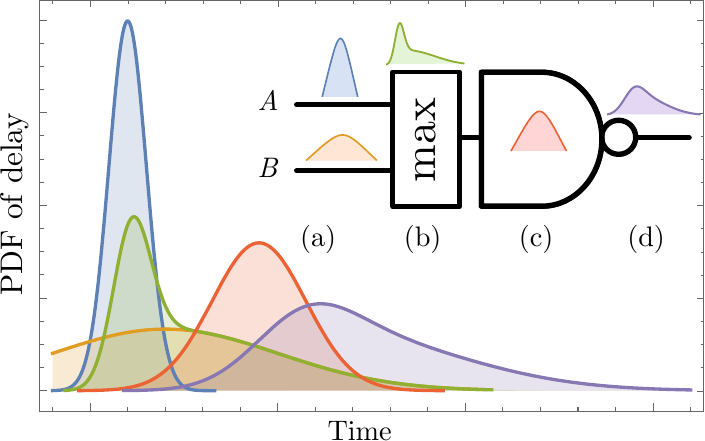}
\caption[Illustration of delay propagation through a logic gate]{Illustration of delay propagation through a logic gate. At stage (a), two signals arrive at the input of the gate. At stage (b), the $\max$-operation is performed, which gives a skewed PDF. At the same time, the gate has its own operation time described by some distribution (c). Thus, the distribution of the gate delay (d) requires the convolution of the obtained distribution (b) and given (c). This convolution results in a new distribution, which clearly has a non-Gaussian form.}
\label{fig:delay_propagation}
\end{figure}

Let us look at the operation of a logic gate in detail. Figure~\ref{fig:delay_propagation} visualises  the step-by-step  transformations of all the PDFs involved into the operation of a logic gates. For the purpose of illustration, we consider that the arrival  and operation time  distributions (\textit{i.e.} the distributions of $X_1$, $X_2$, and $X_0$) are Gaussian. These distributions do not have to be Gaussian, but we want to show that even if \emph{assumed} Gaussian, the convolution~\eqref{eq:convol} will result into a non-Gaussian distribution, as clearly visualised in that Figure. 

Because the PDF for the output of a gate has obviously a non-Gaussian form and does  not imply any analytical solutions, the following ideas have circulated within the Timing Analysis community. The traditional and overall accepted approach is to approximate the $\max$ operation~\cite{blaauw2008StatisticalTimingAnalysis,forzan2009StatisticalStaticTiming}. We have thoroughly discussed different approaches in the review Section~\ref{sec:background}, where we have grouped the block--based approaches into two sub-classes, with phenomenological and parameterised delay models. Azuma~\textit{et al.}~\cite{azuma2017ApproximatingMaximumGaussians} have proposed to model the PDF of the $\max$ with a GMM, which makes it the closest one to the present study. However, the authors did not address the convolution in~\eqref{eq:convol} and did not explain how such a GMM decomposition should be obtained from an arbitrary non-Gaussian distribution.

\begin{figure}[t!]
\center
\includegraphics[width=0.60\textwidth]{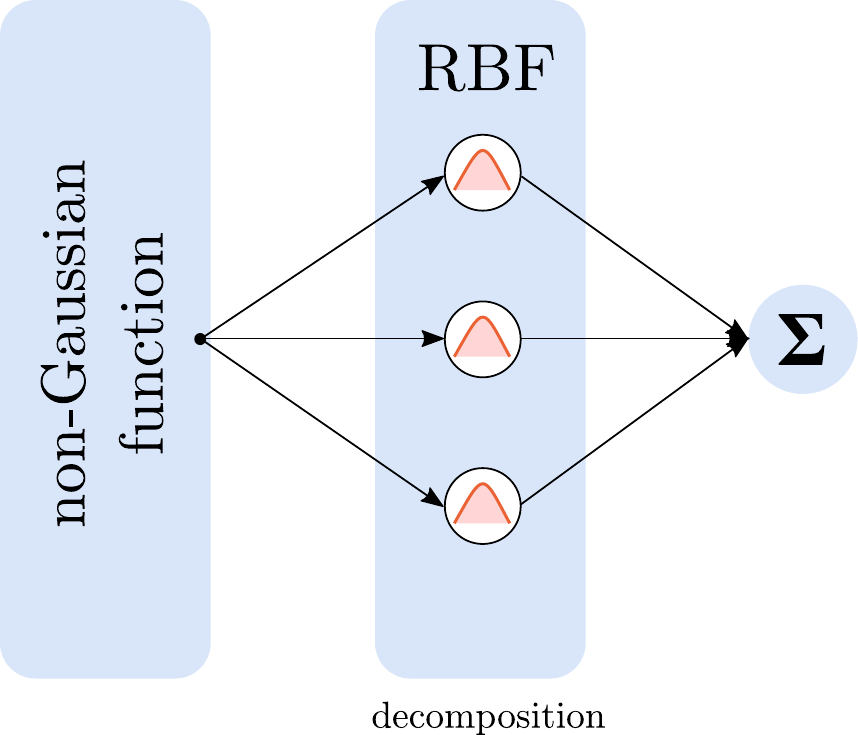}
\caption[RBF representation of a non-Gaussian probability density function]{High-level illustration of the decomposition of a non-Gaussian function and its representation by means of a sum of Radial Basis Functions.}
\label{chap5:fig:function_decompostition}
\end{figure}

In this paper, firstly  we derive the exact analytical formula for the gate delay problem~\eqref{eq:convol} that gives the PDF of~\eqref{eq:gate_operation}. The assumption we make is that the RVs $X_i$ ($i=0,1,2$) in \eqref{eq:gate_operation} are Gaussian ones. Secondly, we propose to handle non-Gaussian distributions as \emph{linear combinations} of functions of the shape \eqref{eq:gauss_pdf}, as schematically shown in Figure~\ref{chap5:fig:function_decompostition}. Such \emph{basis} functions are called RBFs, and this technique is known as GMM and is widely used in signal processing~\cite{kay1993FundamentalsStatisticalSignal} and machine learning~\cite{kelleher2015FundamentalsMachineLearning,murphy2022ProbabilisticMachineLearning}. We shall discuss Gaussian Mixture Models in the next Chapter.
Since the RBFs are Gaussian-like, our approach allows one to traverse a delay's PDF through a timing graph without any loss of information by utilising the exact solution to \eqref{eq:convol}. Note that a common source for such a loss is due to approximation of non-Gaussian PDFs with Gaussian distributions. Algorithms for decomposition a general non-Gaussian distribution into a GMM and further its traversal will be presented in Section~\ref{sec:gakeda}. Now, we will focus on the analytical solution to the gate delay problem~\eqref{eq:convol} and theoretical foundations of our approach.

\section[Gate Delay Distribution]{Distribution of the Gate Delay}
\label{sec:distribution}
\subsection{Maximum of Two Correlated Gaussians} 
Before we move to the analysis, let us make a few useful remarks.  

Firstly, we shall represent symmetric forms like the PDF for the maximum of two correlated Gaussian RVs~\eqref{chap2:eq:max_2_correlated} as follows:
\begin{equation}\label{chap5:eq:max_2_correlated}
  g (x,\rho) = \sum\limits_{\substack{i,j=1,2\\i\neq j}} \phi_i(x) \Phi \left[ \frac{1}{\sqrt{1-\rho^2}} \left( \frac{x-\mu_j}{\sigma_j} -\rho \frac{x-\mu_i}{\sigma_i} \right) \right].
\end{equation}
\begin{figure}[t!]
\center
\includegraphics[width=\textwidth]{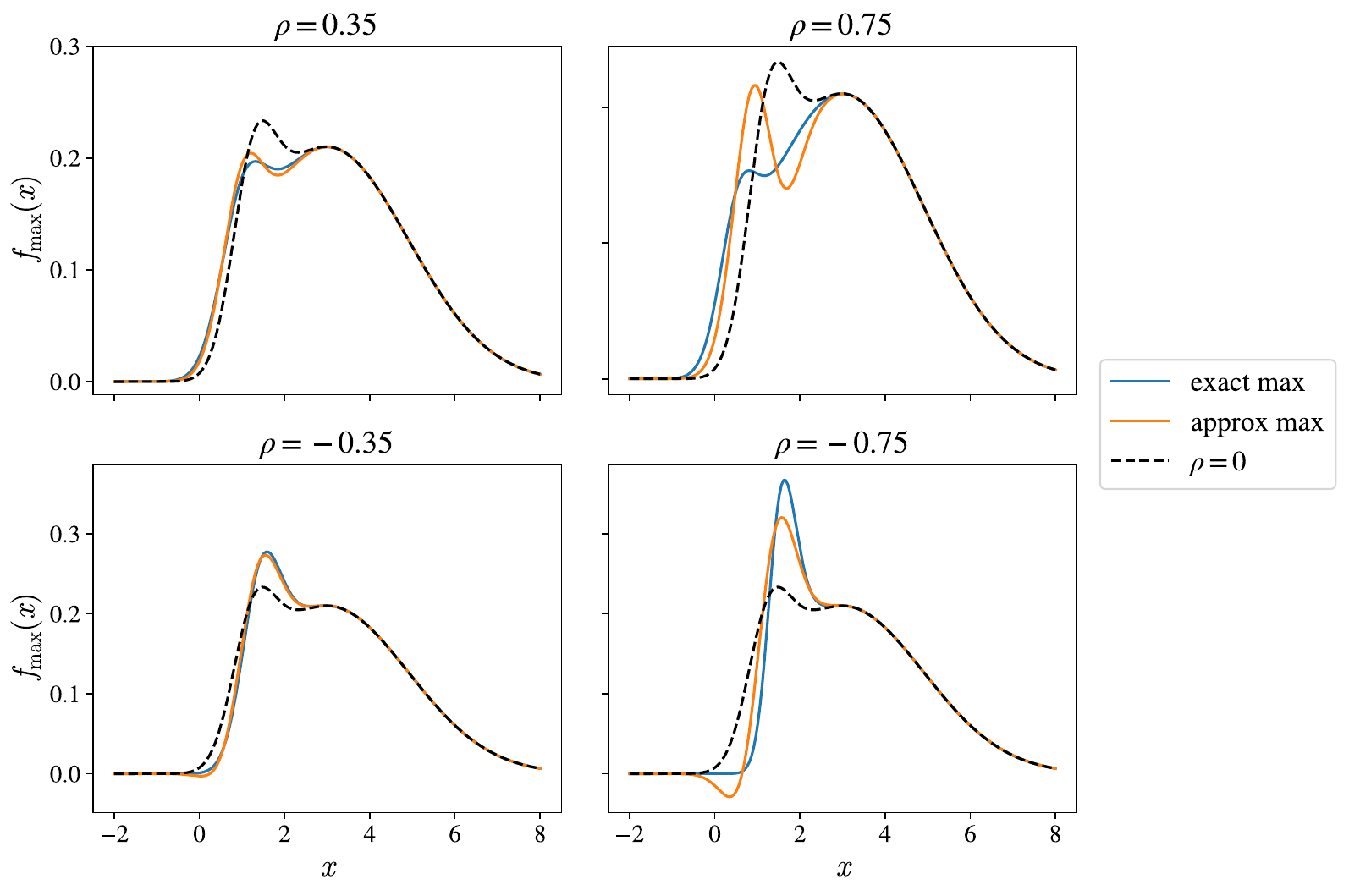}
\caption[PDF of maximum of two Gaussian RVs for weak and strong correlations]{Comparison of different expressions for the PDF of the maximum of two Gaussian RVs, $X_1\sim\mathcal N(1,0.5)$ and $X_2\sim\mathcal N(3,3)$ for different correlation coefficients $\rho$. The blue line shows the exact expression~\eqref{chap5:eq:max_2_correlated}, the orange line shows the  linear-in-$\rho$ expression~\eqref{chap5:eq:max_2_corr_weak}, and the dashed black line corresponds to the independent case~\eqref{chap2:eq:max_2_independent}. For the purpose of illustration, $X_1$ and $X_2$ in this example are two Gaussian RVs with different $\mu$ and $\sigma$.}
\label{chap5:fig:max_weak_corr}
\end{figure}
Note that for $\rho=0$, we immediately obtain the PDF of the maximum of two \emph{independent} Gaussian RVs, given by formula \eqref{chap2:eq:max_2_independent}: $g(x,0) \equiv g(x)$.

Secondly, the key requirement for the algorithms discussed further is \emph{linearity}, as we want to represent delay distributions via sums of RBFs. For this purpose, we consider \emph{weak correlations} between RVs. In the case of a small correlation coefficient, $|\rho|\rightarrow0$, the function $\Phi[\cdot]$ reads
\begin{equation*}
    \Phi \left[ \frac{1}{\sqrt{1-\rho^2}} \left( \frac{x-\mu_j}{\sigma_j} -\rho \frac{x-\mu_i}{\sigma_i} \right) \right]
    = \Phi \left( \frac{x-\mu_j}{\sigma_j} \right) 
    - \phi_j(x) \frac{x-\mu_i}{\sigma_i} \sigma_j \rho + \ldots,
\end{equation*}
and the PDF \eqref{chap5:eq:max_2_correlated} can be approximated in terms of $\rho$ linearly:
\begin{subequations}\label{chap5:eq:max_2_corr_weak}
\begin{equation}\label{eq:max_2_correlated_weak_a}
  g(x,\rho) \approx g(x) + \rho \cdot \delta f_{\text{max}}(x),
\end{equation}
where
\begin{equation}\label{eq:max_2_correlated_weak_b}
  \delta f_{\text{max}}(x) = - \phi_1(x) \phi_2(x) \left[ \frac{x-\mu_1}{\sigma_1} \sigma_2 + \frac{x-\mu_2}{\sigma_2} \sigma_1 \right].
\end{equation}
\end{subequations}

The formulas \eqref{chap5:eq:max_2_correlated} and \eqref{chap5:eq:max_2_corr_weak} are plotted in Figure~\ref{chap5:fig:max_weak_corr} for $X_1\sim\mathcal N(1,0.5)$ and $X_2\sim\mathcal N(3,3)$ and for different correlations coefficients $\rho$. As a reference, the PDF of the maximum or these RVs given by~\eqref{chap2:eq:max_2_independent}, when they are independent, \textit{i.e.} neglecting the correlations, is also shown. As one can see, the linear approximation~\eqref{chap5:eq:max_2_corr_weak} performs well for a small correlation coefficient, however, it breaks down for large $\rho$ as expected.

In the next Section, we derive the PDF for the gate delay distribution~\eqref{eq:convol} assuming initial delays to be Gaussian ones and arrival time delays to be correlated. Then, we will relax this requirement and derive an approximation for weak correlations. We also present the first four moments of the distribution~\eqref{eq:convol}.

\subsection{Derivation of the Exact Formula}
Let us consider three Gaussian RVs, $X_1$, $X_2$ and $X_0$, and an RV $\eta$ such that $\eta = \max(X_1,X_2) + X_0$. We shall assume that RVs $X_1$ and $X_2$ are bounded with the correlation coefficient $\rho$, thus, the PDF of $\max(X_1,X_2)$ is given by \eqref{chap5:eq:max_2_correlated}, and the PDF $\phi_0(x)$ of the RV $X_0$ is given by \eqref{eq:gauss_pdf}. The variables $X_1$ and $X_2$ have the meaning of the arrival times' delays with mean values $\mu_1$, $\mu_2$ and standard deviations $\sigma_1$, $\sigma_2$ correspondingly, \textit{i.e.} $X_1\sim\mathcal N(\mu_1,\sigma_1)$ and $X_2\sim\mathcal N(\mu_2,\sigma_2)$, while $X_0\sim\mathcal N(\mu_0,\sigma_0)$ . Having said that, we can write~\eqref{eq:convol} explicitly implying the Gaussian nature of its components $X_i$:
\begin{equation}\label{eq:convol_2}
  f_{\text{gate}} (x) = \int\limits_{-\infty}^{\infty}g(x',\rho)\phi_0(x-x')dx'.
\end{equation}

Substituting \eqref{chap5:eq:max_2_correlated} into the latter equation, we obtain a sum of two integrals of the form:
\begin{equation*}
  I_{ij}^{(0)} = \int\limits_{-\infty}^{\infty}\phi_i(x') \phi_0(x-x')
  \times \Phi\left[ \frac{1}{\sqrt{1-\rho^2}} \left( \frac{x'-\mu_j}{\sigma_j} -\rho \frac{x'-\mu_i}{\sigma_i} \right) \right]dx'.
\end{equation*}
Recalling the exact form of the PDF $\phi(x)$ and the error function representation of the CDF $\Phi(x)$, formulas \eqref{eq:gauss_pdf} and \eqref{eq:gauss_cdf_1} respectively, and introducing the intermediate variables,
\begin{align*}
    a = \frac12 \frac{\sigma_0^2 + \sigma_i^2}{\sigma_0^2 \sigma_i^2}, \quad b = \frac{\sigma_0^2 \mu_i + \sigma_i^2 (x - \mu_0)}{\sigma_0^2 \sigma_i^2}, \quad
    c = - \frac12 \frac{\sigma_0^2\mu_i^2 + \sigma_i^2 (x-\mu_0)^2}{\sigma_0^2 \sigma_i^2},
  \end{align*}
we arrive to the following expression:
\begin{multline}\label{eq:convolution_i1}
  I_{ij}^{(0)} = \frac{\sqrt{2\pi}}{2} \sigma_j A_0 A_1 A_2 \exp \left( \frac{b^2}{4a} + c \right)
  \\
  \times \int\limits_{-\infty}^{\infty} \exp \left[ -a \left( x'- \frac{b}{2a} \right)^2 \right]
  \\
  \times
  \left\{ 1 + \erf\left[ \frac{1}{\sqrt{2(1-\rho^2)}} \left( \frac{x'-\mu_j}{\sigma_j} -\rho \frac{x'-\mu_i}{\sigma_i} \right) \right] \right\} dx'.
\end{multline}
Here, $A_i = 1/\sqrt{2\pi}\sigma_i$ ($i=0,1,2$) are the normalisation factors corresponding to the distributions $\phi_i(x)$. We present these factors explicitly in order to investigate the most general case of Gaussian-like distributions, which are not necessarily unit-normalised. The first integral in \eqref{eq:convolution_i1} simply gives:
\begin{equation*}
  \int\limits_{-\infty}^{\infty} e^{-a \left( x'- \frac{b}{2a} \right)^2} dx' =  \sqrt{\frac{\pi}{a}}.
\end{equation*}
\begin{figure}[!h]
  \centering
    \includegraphics[width=1\textwidth]{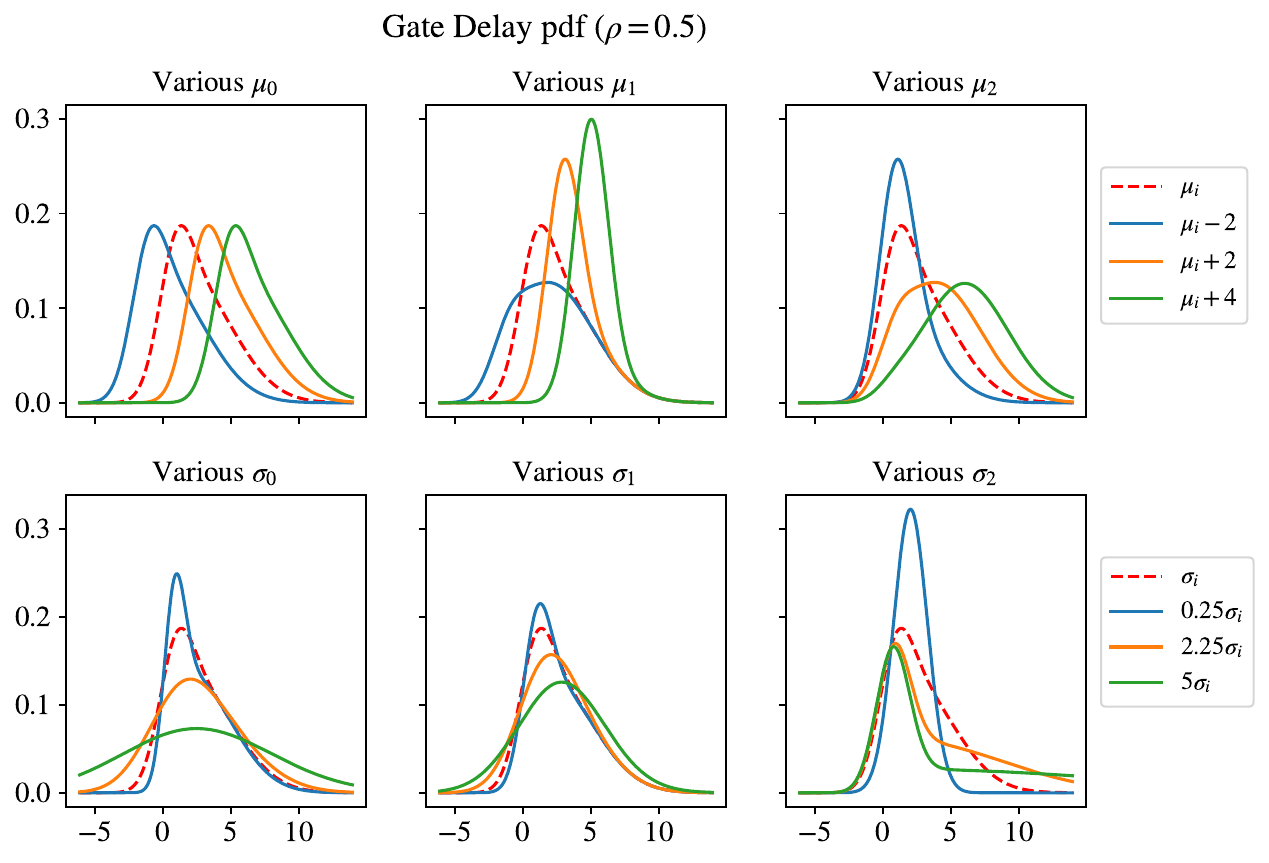}
  \caption[PDF of the gate delay for various parameters]{PDF of the gate delay for a case of $\rho=0.5$. It is taken $X_1\sim\mathcal N(1,0.75)$, $X_2\sim\mathcal N(2,3)$ and $X_0\sim\mathcal N(0,1)$. Then, each of the parameters is varied as specified keeping the rest ones constant.}
  \label{fig:gate_pdf_table}
\end{figure}

The second integral in \eqref{eq:convolution_i1} can be brought to the form given by   formula \eqref{eq:int_gaussian_with_erf} of Appendix~\ref{app:integrals}. Indeed, using a new variable of integration, $t=x'-b/2a$, and introducing
\begin{equation}\label{eq;parameters_k_p}
  \begin{aligned}
    k &= \frac{1}{\sqrt{1-\rho^2}} \left( \frac{1}{\sigma_j} - \frac{\rho}{\sigma_i} \right),
    \\
    p &= \frac{\sigma_0^2\mu_i + \sigma_i^2(x-\mu_0)}{\tilde\sigma_i^2} k - \frac{1}{\sqrt{1-\rho^2}} \left( \frac{\mu_j}{\sigma_j} - \rho\frac{\mu_i}{\sigma_i} \right),
  \end{aligned}
\end{equation}
for the second integral, we have
\begin{equation*}
  \int\limits_{-\infty}^{\infty} e^{-at^2} \erf \left( \frac{k t+p}{\sqrt2} \right) dt 
  = \sqrt{\frac{\pi}{a}} \cdot \erf \left( \frac{p}{\sqrt2} \sqrt{\frac{a}{a+k^2/2}} \right).
\end{equation*}
The integration procedure in the formula above is not trivial and discussed in  Appendix~\ref{app:integrals} mentioned earlier.

Now we can write the formula for the distribution of the logic gate delay. Noting that
\begin{equation*}
  \sqrt{\frac{\pi}{a}} = \sqrt{2\pi} \frac{\sigma_0\sigma_i}{\sqrt{\sigma_0^2 + \sigma_i^2}},
  \quad
  \frac{b^2}{4a} + c = - \frac12 \frac{(x-\mu_0-\mu_i)^2}{\sigma_0^2+\sigma_i^2},
\end{equation*}
we change  to the initial variable of integration instead of the variables $k$ and $p$, and we return to the $\Phi$-representation. After simple algebraic manipulations, we finally obtain:
\begin{subequations}\label{eq:pdf_gate_corr_all}
\begin{multline}\label{eq:pdf_gate_corr_a}
  f_{\text{gate}} (x,\rho) = C\sum\limits_{\substack{i,j=1,2\\i\neq j}}\frac{1}{\tilde\sigma_i} \varphi \left( \frac{x-\mu_0-\mu_i}{\tilde\sigma_i} \right)
  \\
  \times
  \Phi \left[ \frac{1}{ \sqrt{1 + \kappa_{ij}(\rho)^2} }
  \left( \frac{\sigma_i^2 (x - \mu_0) + \sigma_0^2\mu_i}{\tilde\sigma_i\sigma_i\sigma_0} \kappa_{ij}(\rho)
  - \frac{\sigma_i\mu_j - \rho \sigma_j\mu_i}{\sigma_i\sigma_j\sqrt{1-\rho^2}} \right) \right],
\end{multline}
where $C = 2\pi A_0A_1A_2 \sigma_0\sigma_1\sigma_2$ and
\begin{equation}\label{eq:parameter_kappa_rho}
    \begin{aligned}
        \kappa_{ij}(\rho) &= \frac{\sigma_i - \rho\sigma_j}{\sqrt{1-\rho^2}} \cdot \frac{\sigma_0}{\sigma_j \tilde\sigma_i}
  = \frac{\kappa_{ij}}{\sqrt{1-\rho^2}} \left( 1 - \rho \frac{\sigma_j}{\sigma_i} \right),
  \\
  \kappa_{ij} &= \frac{\sigma_i \sigma_0}{\sigma_j \tilde \sigma_i}, \quad
  \tilde\sigma_i = \sqrt{\sigma_0^2 + \sigma_i^2} \,.
  \end{aligned}
\end{equation}
Separating $\kappa_{ij}$ explicitly in \eqref{eq:parameter_kappa_rho}, we can write the formula for the PDF of the logic gate delay as follows:
\begin{multline}\label{eq:pdf_gate_corr_b}
  f_{\text{gate}} (x,\rho) =\,\, C\sum\limits_{\substack{i,j=1,2\\i\neq j}}\frac{1}{\tilde\sigma_i} \varphi \left( \frac{x-\mu_0-\mu_i}{\tilde\sigma_i} \right)
  \\
  \times\Phi \left\{ \left( \frac{1-\rho^2}{ \left( 1-\rho \sigma_j/\sigma_i \right)^2 } + \kappa_{ij}^2 \right)^{-\frac12} \right.
  \\
  \times \left.
  \left[ \eta_{ij}(x) \cdot \mathrm{sign} \left( 1-\rho \frac{\sigma_j}{\sigma_i} \right) - \frac{\mu_j}{\sigma_j} \frac{1-\rho \sigma_j\mu_i/\sigma_i\mu_j}{|1-\rho \sigma_j/\sigma_i|} \right] \right\}.
\end{multline}
For convenience in the formula above, we introduced the following function:
\begin{equation}
\eta_{ij}(x) = \frac{\sigma_i^2 (x - \mu_0) + \sigma_0^2\mu_i}{\tilde\sigma_i^2\sigma_j},
\end{equation}
and the $\mathrm{sign}$-function is defined in a conventional manner:
\begin{equation}
    \mathrm{sign} \left( 1-\rho \frac{\sigma_j}{\sigma_i} \right)
    =
    \begin{cases}
        1, & 1>\rho\sigma_j/\sigma_i;\\
        0, & 1=\rho\sigma_j/\sigma_i;\\
        -1, & 1<\rho\sigma_j/\sigma_i.
    \end{cases}
\end{equation}
\end{subequations}

Let us discuss the obtained result. The formulas \eqref{eq:pdf_gate_corr_all} give an \emph{exact} solution to the gate delay problem~\eqref{eq:convol} assuming that arrival times and gate operation time (interconnect delay, etc.) are given by Gaussian distributions. The resulting distribution \eqref{eq:pdf_gate_corr_all} has a characteristic non-Gaussian form and is fully determined by the parameters of the individual distributions of the signals, \textit{i.e.} by mean values $\mu_i$ and stds $\sigma_i$ of the arrival times and gate operation time. The functions in \eqref{eq:pdf_gate_corr_all}, namely $\varphi(.)$ and $\Phi(.)$, are the Gaussian kernel and Gaussian CDF given by equations \eqref{eq:gauss_pdf} and \eqref{eq:gauss_cdf_1} respectively. These functions are well implemented in standard math environments, which makes computation of \eqref{eq:pdf_gate_corr_all} efficient.

The representations of the gate delay \eqref{eq:pdf_gate_corr_a} and \eqref{eq:pdf_gate_corr_b} are equivalent, and the PDF is shown in Figure~\ref{fig:gate_pdf_table}. However, from formula~\eqref{eq:pdf_gate_corr_b} we can see that the shape of a gate delay significantly simplifies if the input arrival delays (i) have equal standard deviations, $\sigma_1=\sigma_2$, and (ii) are fully correlated, $\rho=1$. If, in addition, the arrival delays have equal mean values, $\mu_1=\mu_2$, the solution simplifies to a trivial case of the convolution of two Gaussian distributions.

\begin{figure}[t]
  \centering{
    \includegraphics[width=0.65\textwidth]{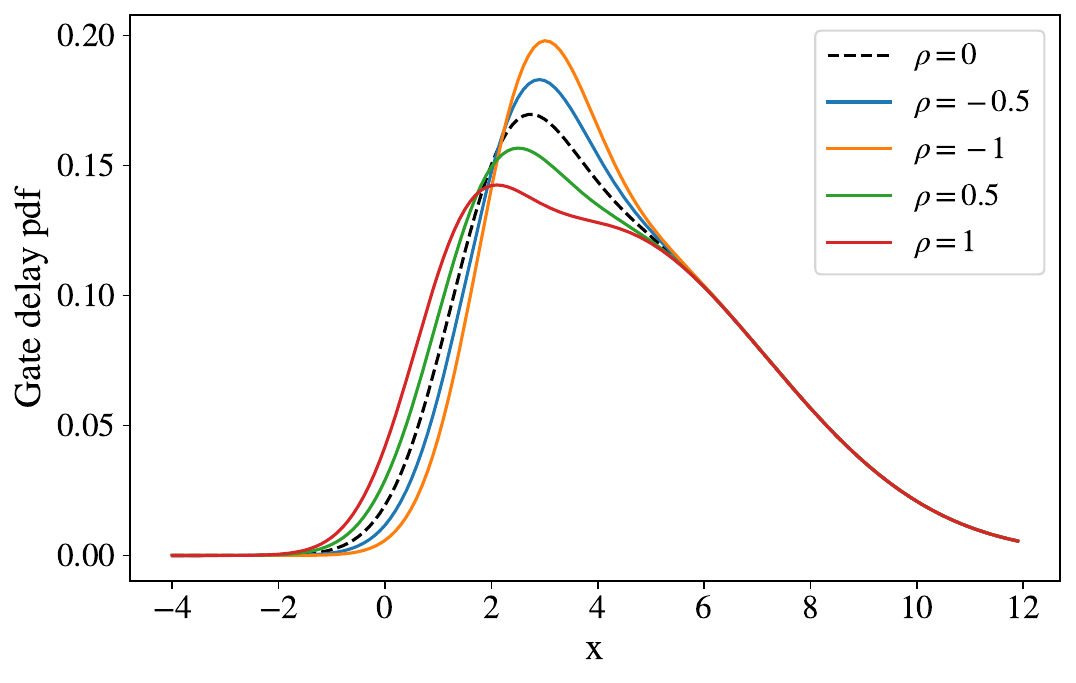}
  }
  \caption[PDF of the gate delay as a function of the correlation coefficient]{A family of curves for the PDF of the gate delay for different values of the correlation coefficient $\rho$. As in Figure~\ref{fig:gate_pdf_table}, the random variables are taken as $X_1\sim\mathcal N(1,0.75)$, $X_2\sim\mathcal N(2,3)$, and $X_0\sim\mathcal N(0,1)$.}
  \label{fig:gate_pdf_corr}
\end{figure}

\subsection{Limit Cases of the Correlation Coefficient}

Let us consider several cases with different strengths of the correlation coefficient $\rho$. The family of curves for various values of $\rho$ is shown in Figure~\ref{fig:gate_pdf_corr}.  The `simplest' result is achieved for independent arrival delays. Hence, for $\rho=0$ the parameter
\begin{equation}
    \kappa_{ij}(\rho=0)=\kappa_{ij} = \frac{\sigma_i \sigma_0}{\sigma_j \tilde \sigma_i},
\end{equation}
and from \eqref{eq:pdf_gate_corr_a} one obtains:
\begin{multline}\label{eq:pdf_gate_independent}
  f_{\text{gate}} (x) = \frac{1}{\sqrt{2\pi}} \sum\limits_{\substack{i,j=1,2\\i\neq j}} \frac{1}{\tilde\sigma_i} \varphi \left( \frac{x-\mu_0-\mu_i}{\tilde\sigma_i} \right)
  \\
  \times\Phi \left[ \frac{1}{\sqrt{1+\kappa_{ij}^2}} \left( \frac{\sigma_i^2 (x - \mu_0) + \sigma_0^2\mu_i}{\tilde\sigma_i^2\sigma_j} - \frac{\mu_j}{\sigma_j} \right) \right],
\end{multline}
where we also have written the value of the coefficient $C$ explicitly.

In the case when correlations are small, $\rho\rightarrow0$, a small parameter in the argument of the function $\Phi\{ \ldots \}$ in \eqref{eq:pdf_gate_corr_a} appears, and we can write:
\begin{multline}
    \Phi \left\{ \ldots \right\} = \Phi \left[ \frac{\eta_{ij}(x) - \mu_j/\sigma_j}{\sqrt{1+\kappa_{ij}^2}} \right]
  \\
  - \frac{1}{\sqrt{2\pi}(1+\kappa_{ij}^2)^{3/2}} \frac{\sigma_j}{\sigma_i}
  \exp \left[ \frac{(\eta_{ij}(x) - \mu_j/\sigma_j)^2}{2(1+\kappa_{ij}^2)} \right]
  \\
  \times
  \left( \eta_{ij}(x) - \frac{\mu_j}{\sigma_j} - \frac{(\mu_i-\mu_j)(1+\kappa_{ij}^2)}{\sigma_j} \right) \rho + \ldots
\end{multline}
Therefore, for the PDF \eqref{eq:pdf_gate_corr_a}--\eqref{eq:pdf_gate_corr_b} in the weak correlation case, we have:
\begin{subequations}\label{chap5:eq:pdf_gate_corr_weak}
\begin{equation}\label{chap5:eq:pdf_gate_corr_weak_a}
    f_{\text{gate}} (x,\rho) \approx f_{\text{gate}} (x) + \rho \cdot \delta f_{\text{gate}} (x),
\end{equation}
where $f_{\text{gate}} (x)$ is the PDF \eqref{eq:pdf_gate_independent} for independent arrival time delays, and the correction $\delta f_{\text{max}}$ reads:
\begin{multline}
    \delta f_{\text{gate}}(x) = -\,\, C\sum\limits_{\substack{i,j=1,2\\i\neq j}}\frac{1}{\tilde\sigma_i} \varphi \left( \frac{x-\mu_0-\mu_i}{\tilde\sigma_i} \right)
    \\
    \times
     \frac{1}{\sqrt{2\pi}(1+\kappa_{ij}^2)^{3/2}} \frac{\sigma_j}{\sigma_i}
     \exp \left[ \frac{(\eta_{ij}(x) - \mu_j/\sigma_j)^2}{2(1+\kappa_{ij}^2)} \right]
     \\
    \times
    \left( \eta_{ij}(x) - \frac{\mu_j}{\sigma_j} - \frac{(\mu_i-\mu_j)(1+\kappa_{ij}^2)}{\sigma_j} \right).
\end{multline}
\end{subequations}

The obtained asymptotic result is \emph{linear} in terms of the correlation coefficient $\rho$, which allows one to use it for the purposes of the algorithm discussed in Section~\ref{sec:gakeda}, where non-Gaussian delays are decomposed into mixtures of RBFs. Here, let us examine the validity of this expansion for different cases. We have already seen, that similar to \eqref{chap5:eq:pdf_gate_corr_weak} linear expression~\eqref{chap5:eq:max_2_corr_weak} for the PDF of the maximum of two Gaussian RVs failed in strong correlation regime (Figure~\ref{chap5:fig:max_weak_corr}). Therefore, let us discuss the issue of \emph{strong} correlations for the approximation $f_{\text{gate}}(x,\rho)$ below.

\begin{figure}[t!]
\center
\includegraphics[width=\textwidth]{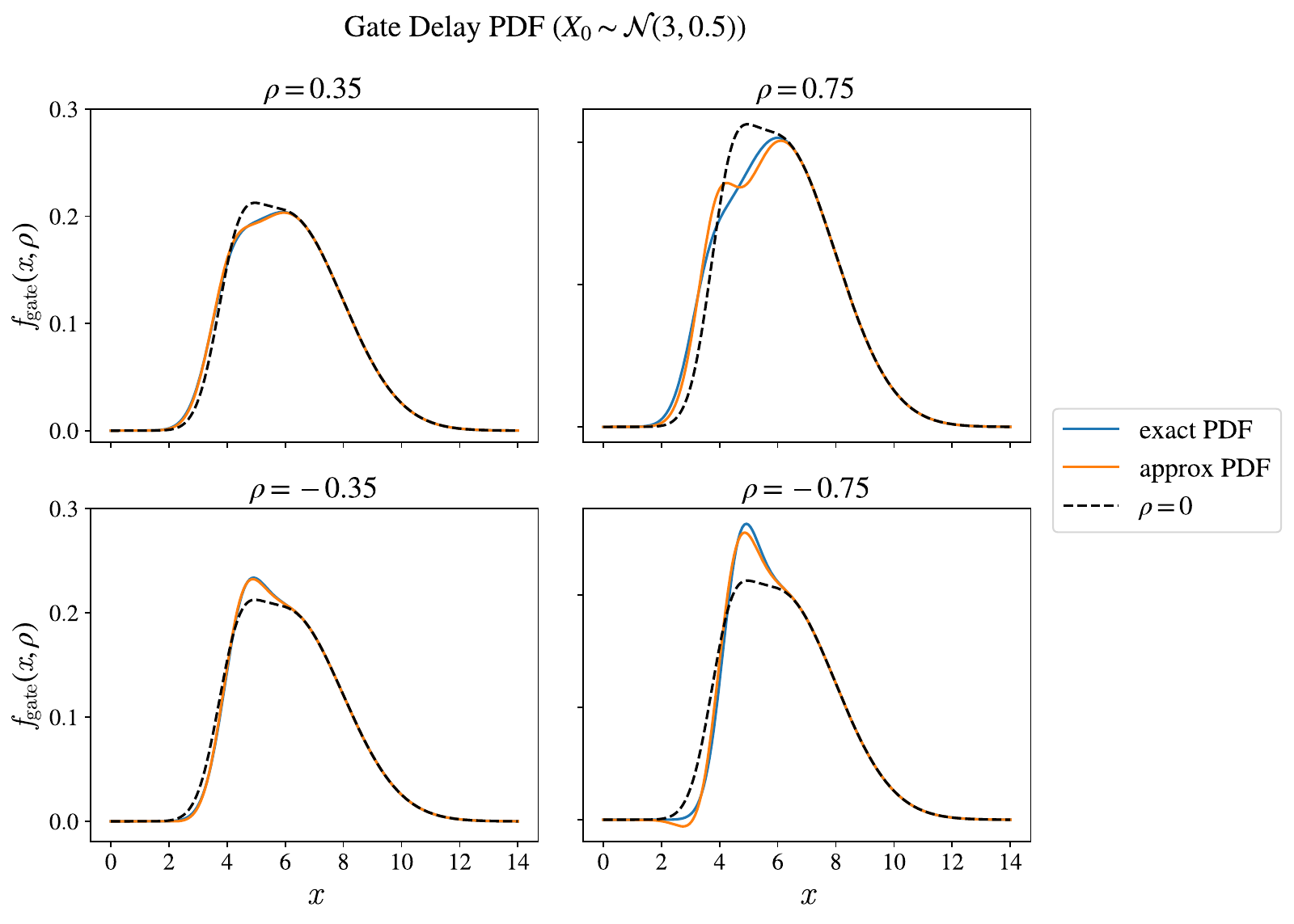}
\caption[Gate delay PDF with weak correlations for $X_0\sim\mathcal N(3,0.5)$]{Comparison of the PDF of the gate delay obtained from different expressions (the case of uncorrelated RVs, approximation and exact expression) for four different correlation coefficients $\rho$. Gate operation time here is $X_0 \sim \mathcal N (3, 0.5)$.}
\label{fig:gate_pdf_corr_X0_3_0.5}
\end{figure}

\begin{figure}[t!]
\center
\includegraphics[width=\textwidth]{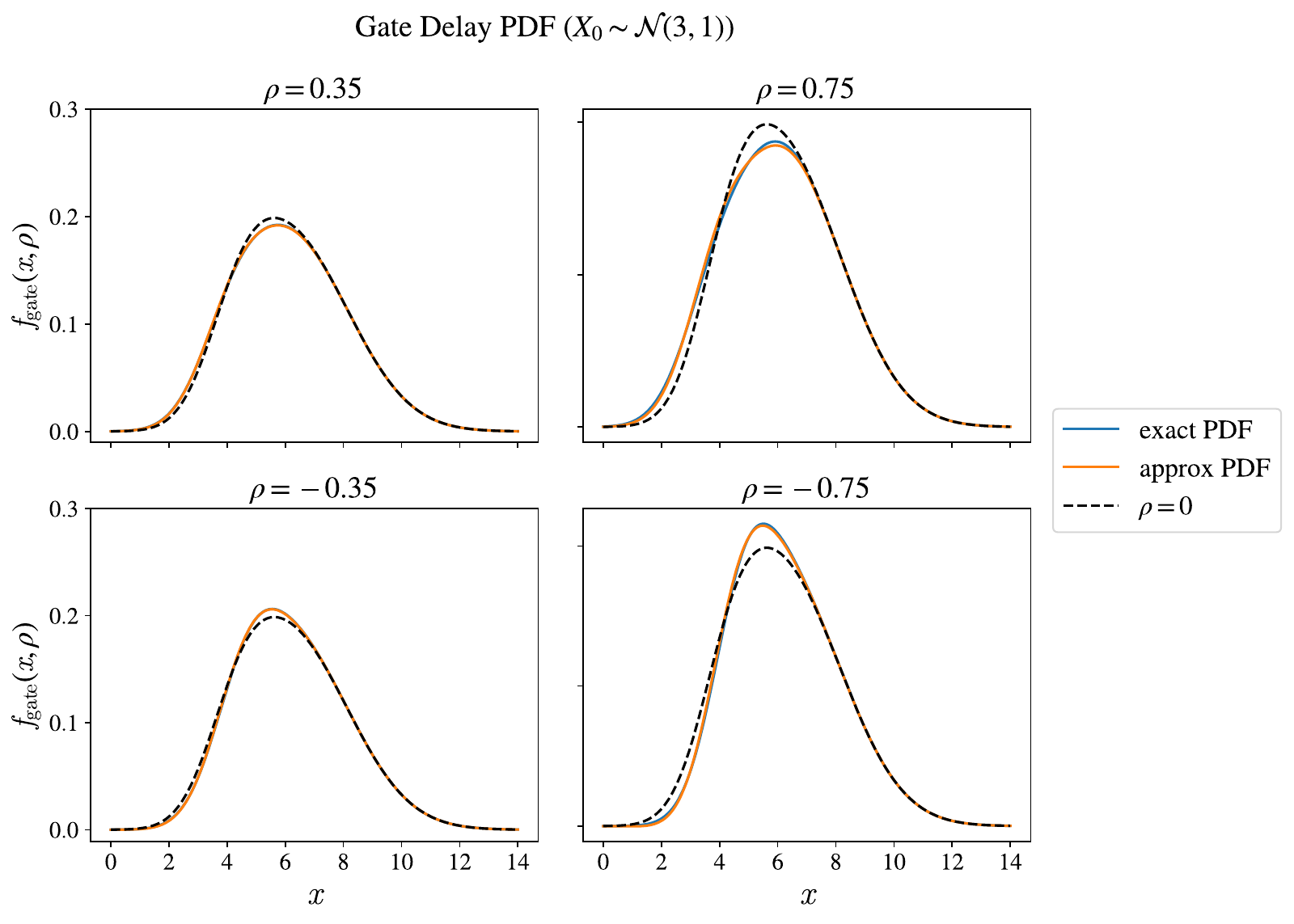}
\caption[Gate delay PDF with weak correlations for $X_0\sim\mathcal N(3,1)$]{Comparison of the PDF of the gate delay obtained from different expressions (the case of uncorrelated RVs, approximation and exact expression) for four different correlation coefficients $\rho$. Gate operation time here is $X_0 \sim \mathcal N (3, 1)$.}
\label{fig:gate_pdf_corr_X0_3_1}
\end{figure}

Figures~\ref{fig:gate_pdf_corr_X0_3_0.5}--\ref{fig:gate_pdf_corr_X0_3_1} show a comparison of the exact result~\eqref{eq:pdf_gate_corr_all} with the linear in $\rho$ approximation~\eqref{chap5:eq:pdf_gate_corr_weak} for different values of the correlation coefficient $\rho$. The random variables $X_{1,2}$ are taken with Gaussian distributions: $X_1\sim\mathcal N(1,0.5)$ and $X_2\sim\mathcal N(3,1.9)$. In Figure~\ref{fig:gate_pdf_corr_X0_3_0.5}, the RV $X_0\sim\mathcal N(3,0.5)$, and we can see that the linear approximation also fails when the correlations between the inputs are strong. At the same time, Figure~\ref{fig:gate_pdf_corr_X0_3_1} shows the case with $X_0\sim\mathcal N(3,1)$, \textit{i.e.} with the twice larger standard deviation, and the linear approximation~\eqref{chap5:eq:pdf_gate_corr_weak} gives perfect agreement with the exact solution.

This should not be surprising, if one recalls that the gate delay PDF is the \emph{convolution} of $g(x,\rho)$ with the Gaussian PDF $\phi_0(x)$, which describes the delay due to the gate operation time, interconnect delay, \emph{etc}. Thus, when the variance  of the latter is high, its contribution in the resulting delay~\eqref{eq:convol_2} becomes dominating, and inaccuracies in the approximation of $g(x,\rho)$ vanish.

\section{Moments of the Gate Delay Distribution}

In this Section, we consider the first moments of the distribution \eqref{eq:pdf_gate_corr_a}. In general, the moments are given by the following expressions:
\begin{equation}\label{eq:gate_moments_general}
  \mathbb E[X^n] = \int\limits_{-\infty}^{\infty} x^n f_{\text{gate}}(x,\rho) dx.
\end{equation}
Below, we discuss the first four moments: the mean, standard deviation, skewness and kurtosis of the gate delay PDF. The meaning of these moments is explained in Section~\ref{chap2:sec:basic_prob}.

\subsection{Mean}
\begin{figure}[t!]
\center
\includegraphics[width=0.8\textwidth]{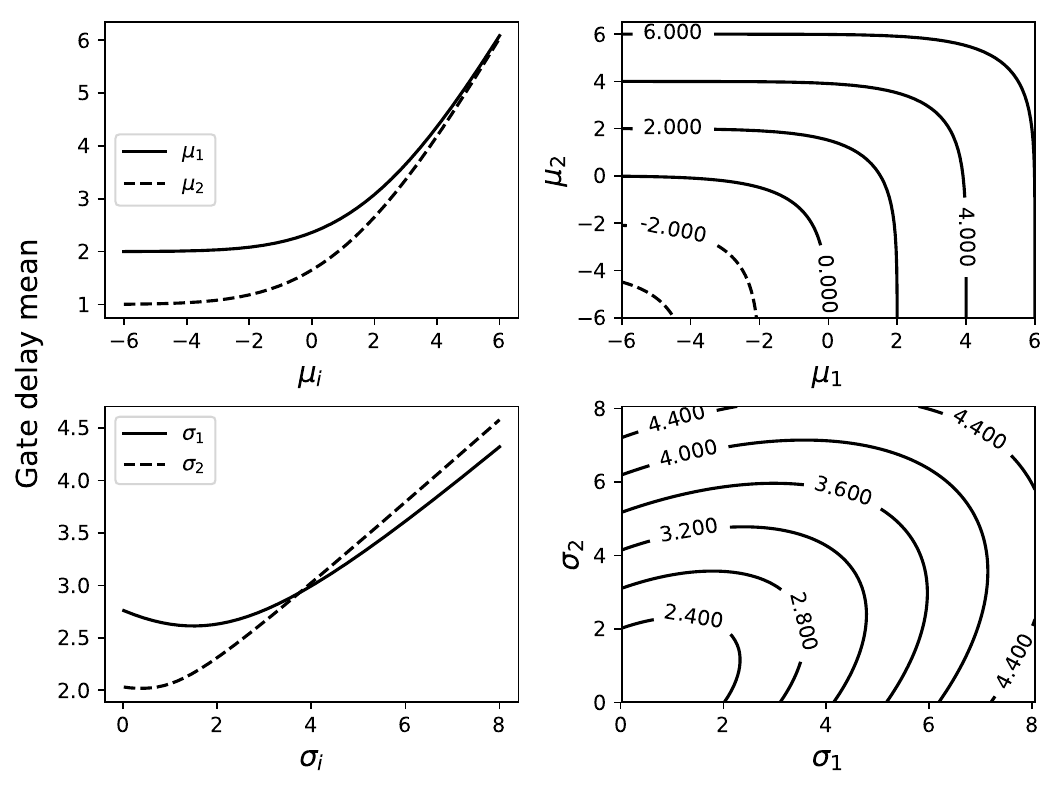}
\caption[Gate delay mean value]{Mean value of the gate delay as a function of the input delay mean values and standard deviations for $\rho=0.5$ and $X_0\sim\mathcal N(0,1)$.}
\label{fig:gate_mean_table}
\end{figure}

The mean of the gate delay is given by
\begin{equation}\label{eq:gate_mean_general}
  \mu_{\text{gate}}\equiv\mathbb E[X] = \int\limits_{-\infty}^{\infty} x f_{\text{gate}}(x,\rho) dx.
\end{equation}
The substitution of \eqref{eq:pdf_gate_corr_a} into \eqref{eq:gate_mean_general} leads to the calculations, similar to those performed above. Therefore, in this Section we only present the final results for the moments. The mean value of the gate delay reads:
\begin{multline}\label{eq:gate_mean}
    \mu_{\text{gate}} = C \sum\limits_{\substack{i,j=1,2\\i\neq j}}
    \left\{ \sqrt{2\pi}(\mu_0+\mu_i) \Phi\left[ \frac{1}{\sqrt{1-\rho^2}} \frac{\mu_i-\mu_j}{\sigma_j\varkappa_{ij}} \right] \right.
    \\
    + \left. \frac{\sigma_i\tilde\sigma_i}{\sigma_0} \frac{\kappa_{ij}(\rho)}{\varkappa_{ij}} \exp \left[ - \frac12 \frac{1}{1-\rho^2} \frac{(\mu_i-\mu_j)^2}{\sigma_j^2 \varkappa_{ij}^2 } \right] \right\},
\end{multline}
where
\begin{equation}\label{eq:varkappa}
  \varkappa_{ij} = \sqrt{1+\kappa_{ij}(\rho)^2\left( 1 + \frac{\sigma_i^2}{\sigma_0^2} \right)}.
\end{equation}
Figure~\ref{fig:gate_mean_table} shows the dependencies of the mean value $\mu_{\text{gate}}$ on the mean values of input delays and the standard deviations of input delays. The gate operation time is taken as a standard Gaussian $X_0\sim\mathcal N(0,1)$, and the correlation coefficient is $\rho=0.5$.

\subsection{Standard Deviation}
The evaluation of the standard deviation requires the knowledge of the second central moment:
\begin{equation}\label{eq:gate_std_general}
    \sigma_{\text{gate}} \equiv \mathbf D X \equiv \sqrt{\langle X^2 \rangle - \langle X \rangle^2} = \sqrt{\mathbb E(X^2) - \mathbb E(X)^2}.
\end{equation}
Performing the integration in \eqref{eq:gate_moments_general} for $n=2$, after some algebra, for the second central moment one obtains:
\begin{multline}
  \mathbb E[X^2] = C \sum\limits_{\substack{i,j=1,2\\i\neq j}}
  \left\{ \sqrt{2\pi} \left[ (\mu_0+\mu_i)^2 + \tilde\sigma_i^2 \right] \Phi\left[ \frac{y(\ldots)}{\varkappa_{ij}} \right] \right. 
    \\
    + 2\frac{\sigma_i\tilde\sigma_i^2}{\sigma_0} \frac{\kappa_{ij}(\rho)}{\varkappa_{ij}} \left( \frac{\mu_0+\mu_i}{\tilde\sigma_i} - \frac12 \frac{\sigma_i}{\sigma_0} \frac{\kappa_{ij}(\rho)}{\varkappa_{ij}^2} y(\ldots) \right) 
    \\
    \times \left. \exp \left[ - \frac12 \frac{y(\ldots)^2}{\varkappa_{ij}^2} \right] \right\},
\end{multline}
where
\begin{equation}
  y(\ldots) = \frac{\tilde\sigma_i \mu_i}{\sigma_i\sigma_0} \kappa_{ij}(\rho) - \frac{\mu_j}{\sigma_j} \frac{1-\rho \sigma_j\mu_i/\sigma_i\mu_j}{\sqrt{1-\rho^2}}\,.
\end{equation}

\begin{figure}[t!]
\center
\includegraphics[width=0.8\textwidth]{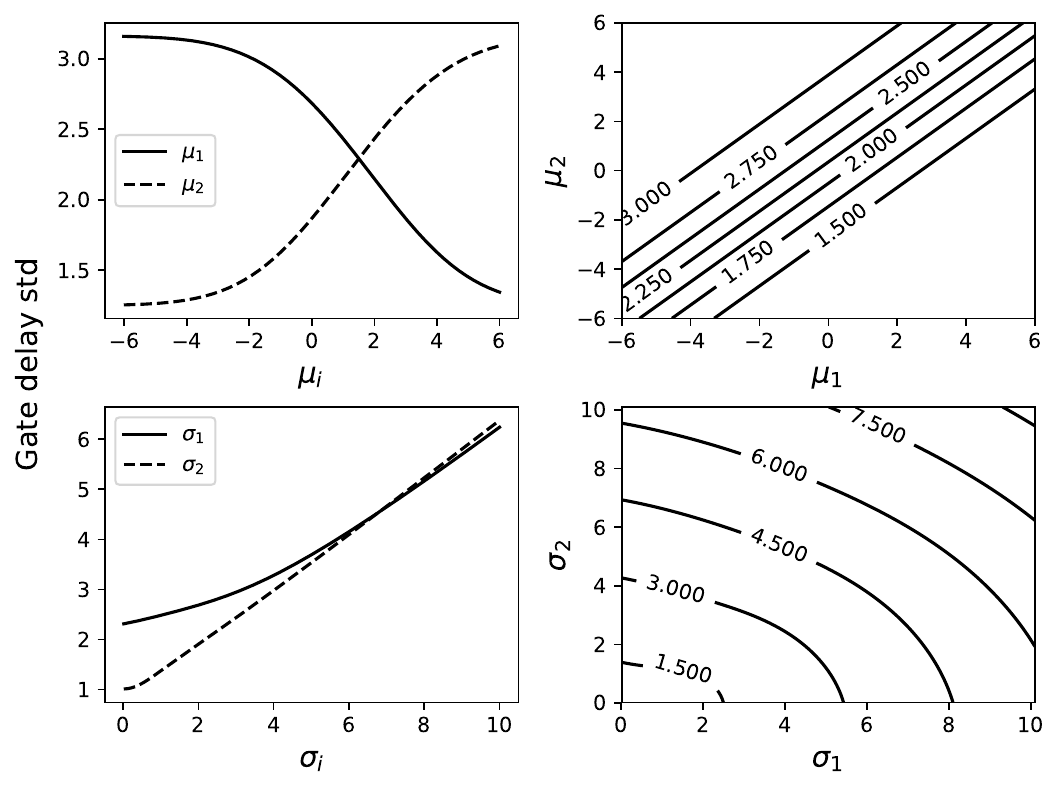}
\caption[Gate delay standard deviation]{Gate delay standard deviation as a function of $\mu_{1,2}$ and $\sigma_{1,2}$. The parameters are the same as in Figure~\ref{fig:gate_mean_table}.}
\label{fig:gate_std_table}
\end{figure}

Now, having both first and second central moments, $\mathbb E[X]$ and $\mathbb E[X^2]$, we can calculate the standard deviation $\sigma_{\text{gate}}$ of the gate delay. This quantity is presented in Figure~\ref{fig:gate_std_table} for the same values of the parameters as in the Figure with  the mean value $\mu_{\text{gate}}$.


\subsection{Higher Moments}
\begin{figure}[t!]
\center
\includegraphics[width=\textwidth]{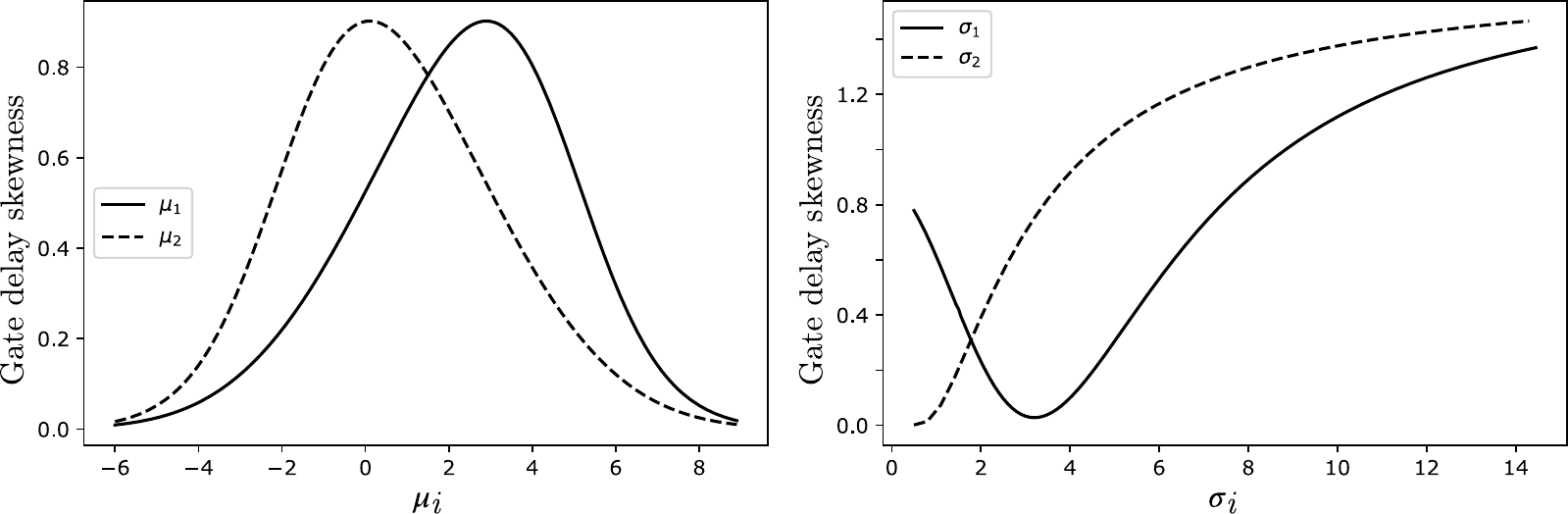}
\caption[Gate delay skewness]{Gate delay skewness as the function of input delays' mean values and standard deviations for $\rho=0.5$ and $X_0\sim\mathcal N(0,1)$.}
\label{fig:gate_delay_skewness}
\end{figure}

\begin{figure}[t!]
\center
\includegraphics[width=\textwidth]{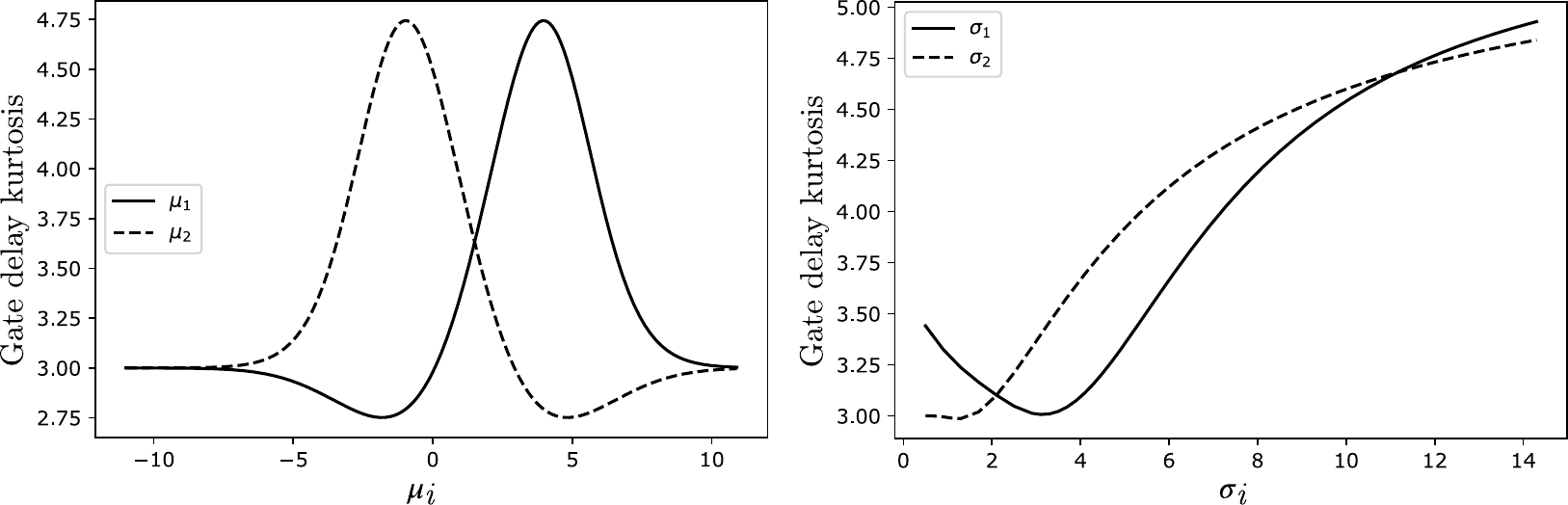}
\caption[Gate delay kurtosis]{Gate delay kurtosis as the function of input delays' mean values and standard deviations for $\rho=0.5$ and $X_0\sim\mathcal N(0,1)$.}
\label{fig:gate_delay_kurtosis}
\end{figure}

The higher moments can be obtained in the same manner. We do not present the exact expressions due to their cumbersomeness. However, gate delay skewness
\begin{equation}\label{eq:gate_skewness}
  \gamma_{\text{gate}} \equiv \operatorname{E}\left[\left(\frac{X-\mu}{\sigma}\right)^3 \right] = \frac{\operatorname{E}[X^3] - 3\mu_{\text{gate}}\sigma_{\text{gate}}^2 - \mu_{\text{gate}}^3}{\sigma_{\text{gate}}^3},
\end{equation}
and the kurtosis
\begin{equation}\label{eq:gate_kurtosis}
  \kappa_{\text{gate}} \equiv \operatorname{E}\left[\left(\frac{X-\mu}{\sigma}\right)^4 \right]
     = \frac{\operatorname{E}[(X-\mu)^4]}{\sigma^4},
\end{equation}
are shown in Figures~\ref{fig:gate_delay_skewness}--\ref{fig:gate_delay_kurtosis} correspondingly.

\section{Delay Propagation Algorithm}
\label{sec:gakeda}
\subsection{Brief Theoretical Summary}

Consider a logic gate with two inputs, $A$ and $B$, and suppose that the gate operation time is distributed according to the normal law with a mean $\mu_0$ and a variance $\sigma_0^2$, \textit{i.e.} it is a Gaussian RV. Assume now that the arrival times of both signals are also Gaussian RVs with means and variances ($\mu_1$, $\sigma_1^2$) and ($\mu_2$, $\sigma_2^2$) respectively. Even if the individual distribution in the example of Figure~\ref{fig:delay_propagation} are Gaussian, the application of formula~\eqref{eq:gate_operation} leads to a non-Gaussian distribution of the delay at the gate output as shown in that figure.
The exact expression for the PDF of such an RV is
\begin{equation}\label{chap6:eq:pdf_gate}
  f_{\text{gate}}(x) = \frac{1}{\sqrt{2\pi}} \sum\limits_{\substack{i,j=1,2\\i\neq j}} \frac{1}{\tilde\sigma_i} \varphi \left( \frac{x-\mu_0-\mu_i}{\tilde\sigma_i} \right)
  \Phi \left[ \frac{1}{\sqrt{1+\kappa_{ij}^2}} y(\ldots) \right],
\end{equation}
where
\begin{equation}\label{eq:parameter_kappa}
  \kappa_{ij} = \frac{\sigma_0 \sigma_i}{\sigma_j \tilde\sigma_i}, \quad \tilde\sigma_i = \sqrt{\sigma_0^2 + \sigma_i^2},
  \quad
  y(x) = \frac{\sigma_i^2 (x - \mu_0) + \sigma_0^2\mu_i}{\tilde\sigma_i^2\sigma_j} - \frac{\mu_j}{\sigma_j}.
\end{equation}

While expression~\eqref{chap6:eq:pdf_gate} does not take into account possible correlations between the arrival signals, we are interested in \emph{demonstrating} how this exact solution can speed up SSTA for a given graph keeping precision high. The full expression that takes the correlations into account was derived above and is given by equation~\eqref{eq:pdf_gate_corr_all}. Formula~\eqref{chap6:eq:pdf_gate} assumes all initial delays (arrival and gate itself) to have Gaussian distributions. In principle, both the arrival signal and gate delay do not have to be Gaussian. If they can be decomposed into a linear superposition of Gaussian kernel functions, the PDF of the gate output delay can be presented as a linear combination of expressions~\eqref{chap6:eq:pdf_gate} due to the linearity of the integration operation. This idea constitutes the core of a delay propagation algorithm which we discuss in the next Sections. 

\subsection{Algorithm}

The algorithm \texttt{GaKeDA} for the calculation of the delay propagation through a timing graph is shown in Figure~\ref{fig:gakeda_algo} and outlined below. The key feature of the algorithm is representing non-Gaussian distributions with a mixture of RBFs that have Gaussian form, \textit{i.e.} as a linear combination (weighted sum) of Gaussian kernels $\varphi(x)$ given by \eqref{eq:gauss_pdf}. This allows one to use the result \eqref{chap6:eq:pdf_gate}, obtained above for Gaussian distributions. Hence the name for the algorithm: \textbf{Ga}ussian \textbf{Ke}rnel \textbf{D}ensity Estimation based \textbf{A}lgorithm (\texttt{GaKeDA}). The Algorithm returns a \texttt{list} with the parameters for the Gaussian mixtures for each node of the graph $G$.

\begin{figure}[!t]
\center
  \begin{algorithm}[H]
  \KwData{graph $G$ and delay distributions for nodes and input signals}
  \KwResult{a list with PDFs for all nodes in the RBF representation}
  \nl\tcc{We assume that the sequence, in which the nodes should be visited, is known}
  \For{node in $G$}
  {
    $f_{node} \gets$ initialise\;
    \If{delay's PDF not Gaussian}
        {
            $f_{component} \gets$ utilise the exact formula \eqref{chap6:eq:pdf_gate}\; 
            $f_{node} \gets f_{node} + f_{component}$\;
      }
    \Else{
        $f_{node} \gets $ utilise the exact formula \eqref{chap6:eq:pdf_gate}; \Comment{delay is Gaussian}
    }
    $f_{GMM} \gets $ decompose $f_{node}$ into GMM \;
    add GMM parameters to a list
  }
  \end{algorithm}
\caption[\texttt{GaKeDA} Algorithm]{\texttt{GaKeDA} Algorithm.}
\label{fig:gakeda_algo}
\end{figure} 

Such models that use Gaussian kernels \eqref{eq:gauss_pdf} as the shape functions are called GMMs and are well known and studied in the literature
There are numerous books and monographs on mixture models, not only GMMs, that cover theoretical foundations and applications, \textit{e.g.} by Everitt~\& Hand~\cite{everitt1981FiniteMixtureDistributions}, Titterington~\textit{et al.}~\cite{titterington1985StatisticalAnalysisFinite}, McLachlan~\& Basford \cite{mclachlan1987MixtureModelsInference}, Lindsay~\cite{lindsay1995MixtureModelsTheory}, B\"ohning~\cite{bohning1999ComputerassistedAnalysisMixtures}, McLachlan~\& Peel~\cite{mclachlan2000FiniteMixtureModels}, Fr\"uhwirth--Schnatter~\cite{fruhwirth-schnatter2006FiniteMixtureMarkov}, Mengersen~\textit{et al.}~\cite{mengersen2011MixturesEstimationApplications}, and McNicholas~\cite{mcnicholas2016MixtureModelBasedClassification}.
A~recent review by McLachlan~\textit{et al.}~\cite{mclachlan2019FiniteMixtureModels} was published in 2019. Applications of mixture models in Machine Learning are covered in a book by Murphy~\cite{murphy2022ProbabilisticMachineLearning}. GMMs are realised in various software packages such as \texttt{scikit-learn}~\cite{pedregosa2011ScikitlearnMachineLearning}.

In case the arrival signals' delays and operation time of a gate have Gaussian distribution, the formula \eqref{chap6:eq:pdf_gate} is applied straightforwardly (see the line $9$ of \texttt{GaKeDA} in Figure~\ref{fig:gakeda_algo}). When one of the delays has non-Gaussian distribution, it should be first decomposed into a Gaussian mixture and, therefore, the formula~\eqref{chap6:eq:pdf_gate} is applied for every pair of the mixture components (see the lines $5$\,--\,$7$ of \texttt{GaKeDA}). From here, one can see the complexity of the Algorithm: in worst case, when all the delays in $G$ are represented \textit{via} Gaussian mixtures, the complexity with respect to the total number $N$ of the atomic operations is $\mathcal O(m^3N)$, where $m$ is a number of components in the Gaussian mixture. However, the gate operation time often has Gaussian distribution~\cite{champac2018TimingPerformanceNanometer}. In this case, only the arrival times should be represented by means of the mixtures and, thus, the complexity reduces to $\mathcal O(m^2N)$. Such complexity is comparable with that of the known approaches within the block--based SSTA, which we have discussed in Section~\ref{sec:background} (see Tables~\ref{chap3:tab:compexity_a} and~\ref{chap3:tab:compexity_b})

\begin{figure}[t!]
\center
\includegraphics[width=\textwidth]{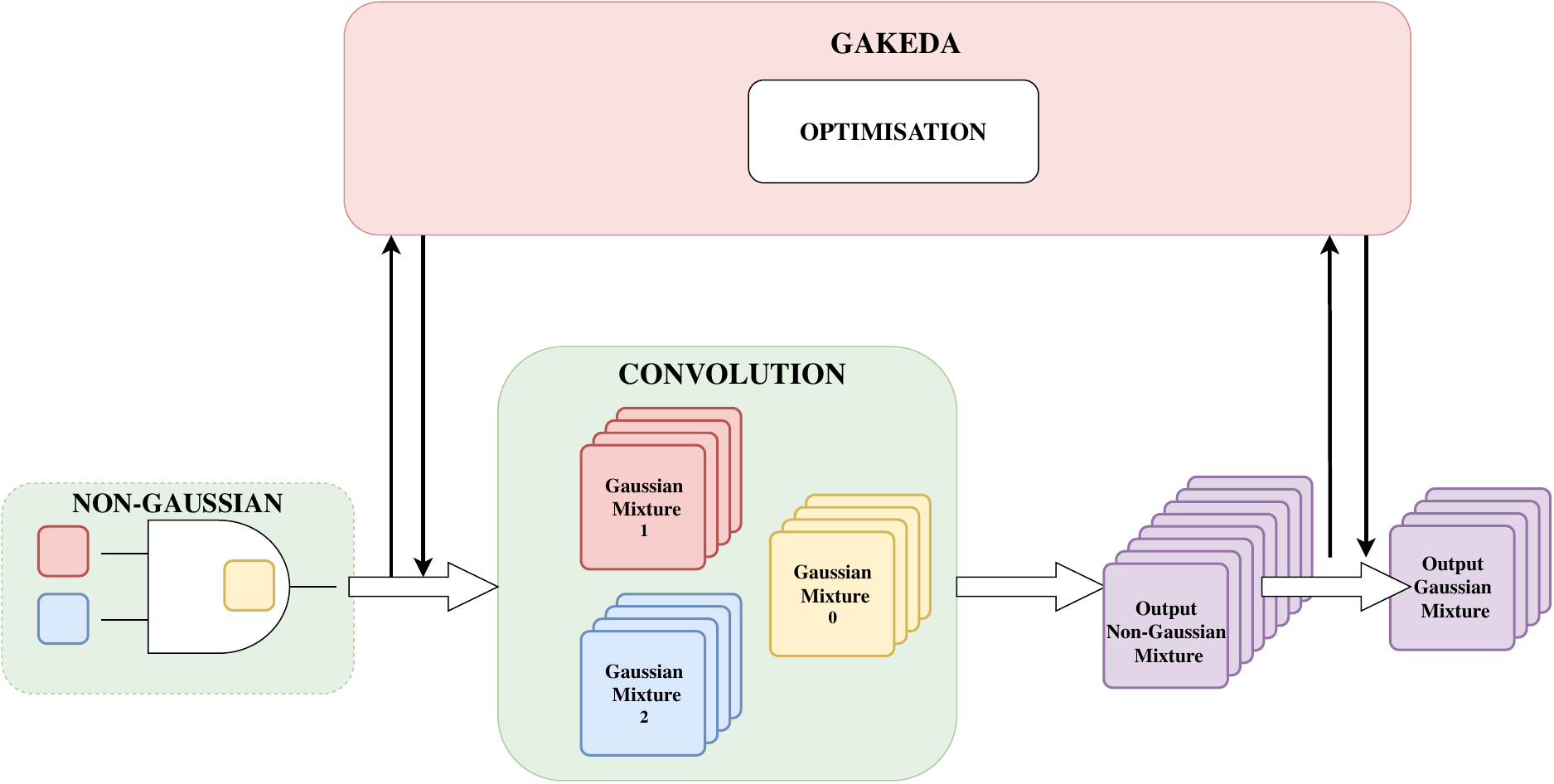}
\caption[High-level diagram of the \texttt{GaKeDA} flow]{High-level diagram of the \texttt{GaKeDA} flow.}
\label{fig:decomposition_flow}
\end{figure}

The high-level diagram of the \texttt{GaKeDA} flow is shown in Figure~\ref{fig:decomposition_flow}. The algorithm relies on the decomposition procedure (see the line $10$ of \texttt{GaKeDA}) to find the parameters of GMM representation. The GMM parameters identification problem is a non-convex optimisation problem, thus, \texttt{GaKeDA} may be inefficient. In the next section, we propose a \textit{Gaussian comb} model for GMM, which allows efficient  GMM parameters identification.

\subsection{Optimisation}

The exact function \eqref{chap6:eq:pdf_gate} can be written as $f_{\text{rbf}}(x)$, a sum of RBFs, and each of these RBFs has a Gaussian-like shape. In other words, it can be decomposed into a Gaussian mixture~\cite{gregor1969AlgorithmDecompositionDistribution,titterington1985StatisticalAnalysisFinite,mclachlan2000FiniteMixtureModels,xu2018ParameterEstimationGaussian}. The decomposition procedure is equivalent to fitting the actual PDF with a sum of RBFs, which brings us to an optimisation problem. In this study, we discuss the minimisation of the sum of squares of the residuals
\begin{equation}\label{eq:least_squares}
    \min \sum_i | f_{\text{rbf}}(x_i) - y_i |^2,
\end{equation}
subject to constraints (specified below). Here $y_i$ are the data points corresponding to the actual function $f_{\text{gate}}(x)$ that we want to fit.

Depending on the form of the RBFs, the minimisation of \eqref{eq:least_squares} can vary significantly, \textit{e.g.} an approximate function $f_{\text{rbf}}(x)$ can be either linearly or non-linearly dependent on the fitting parameters. Different approaches to this problem were discussed by Mishagli \& Blokhina~\cite{mishagli2020RBFApproximationNonGaussian}.

The optimisation problem should be formulated as a LP problem in order to avoid a numerical approximate optimisation procedure and to speed up the algorithm~\cite{boyd2004ConvexOptimization}. Instead of the $2-$norm $||\ldots||_2$, consider the infinite norm
\begin{equation}
    ||\mathbb M \mathbf x - \mathbf y||_{\infty} = \max (|r_1|,|r_2|,\ldots,|r_n|),
\end{equation}
where $|r_i| = |M_i^T \mathbf x - y_i|$ ($i=1,\ldots,n$) are residuals. Denoting $\max (\ldots)$ by $t$, we have the following LP problem:
\begin{equation}
  \begin{array}{ll}
      \min & t \\
      \text{s.t.} & -t \mathbf 1 \preceq \mathbb M \mathbf x - \mathbf y \preceq t \mathbf 1,
      \quad \mathbf x \succeq \mathbf 0.
  \end{array} 
\end{equation}

\noindent Thus, for an LP problem of the form
\begin{equation}
  \begin{array}{ll}
    \min & \mathbf c^T \mathbf x \\
    \text{s.t.} & \mathbb A \mathbf x \preceq \mathbf b,
    \quad \mathbf x \succeq \mathbf 0,
\end{array}
\end{equation}
the vectors $\mathbf c$, $\mathbf x$ and $\mathbf b$ and the matrix $\mathbf A$ are as follows. Having $n$ data points ($x_i,y_i$) and considering $m$ RBFs, the vectors $\mathbf c$ and $\mathbf x$ are
\begin{equation}
    \mathbf c^T = (\underbrace{0,\ldots,0}_{m\text{ times}},1)
    \quad
    \mathbf x^T = (w_1, w_2, \ldots, w_m, t),
\end{equation}
where $w_i$ ($i=1,\dots,m$) are unknown weights.

The matrix $\mathbb A$ of size $(2n,m+1)$ reads
\begin{equation}
    \mathbb A =
    \begin{pmatrix}
        \mathbb M  & -\mathbf 1 \\
        -\mathbb M & -\mathbf 1
    \end{pmatrix},
\end{equation}
where
\begin{equation}
    \mathbb M =
    \begin{pmatrix}
        \varphi_1(x_1) & \varphi_2(x_1) & \ldots & \varphi_m(x_1) \\
        \varphi_1(x_2) & \varphi_2(x_2) & \ldots & \varphi_m(x_2) \\
        \vdots         & \vdots         & \ddots & \vdots \\
        \varphi_1(x_n) & \varphi_2(x_n) & \ldots & \varphi_m(x_n)
    \end{pmatrix}. 
\end{equation}
The vector $\mathbf b^T = (\mathbf y, -\mathbf y)$ is of length $2n$.

\section{Discussion and Conclusions}
\label{sec:discussion}

In this paper, original research contributions in the context of block--based SSTA were presented. The SSTA problem of a logic circuit delay computation was considered at the gate level. A fundamental (mathematical) operation that describes the operation of a logic gate, \textit{i.e.} the \emph{convolution}~\eqref{eq:convol}, was studied. To summarise, the main results of the paper are as follows.
\begin{itemize}
    \item The exact analytical solution to~\eqref{eq:convol} has been given in form of \eqref{eq:pdf_gate_corr_all}. The only assumption made was that the arrival times delays and gate operation times were normally distributed. The expression for the gate delay PDF \eqref{eq:pdf_gate_corr_all} is a function of input distributions parameters, \textit{i.e.} mean values, stds, and the correlation coefficient of the arrival times; it has no unknown fitting parameters. It should be noted that even with the assumption of normally distributed input parameters, such a solution was not known in the field.
    \item The obtained expression for the gate delay has the advantage to the known results, such as by Azuma~\textit{et al.}~\cite{azuma2017ApproximatingMaximumGaussians}, as the convolution is also taken into account, thus, the gate delay problem is fully analysed.
    \item The obtained solution has been studied in both zero and weak correlations cases. In the latter case, the PDF of the gate delay is linear in the correlation coefficient $\rho$. It has also been shown that, depending on the standard deviation of gates operation times, this approximation can be applicable in the strong correlation regime as well.
    \item For the derived distribution, the first and second moments, mean and standard deviation, as well as the higher moments, skewness and kurtosis, have been obtained. 
    \item The theoretical foundations for algorithms that rely on \emph{decomposition} of non-Gaussian PDF into a superposition of RBFs have been given.
    \item The exact formula for an output logic gate delay, the convolution of $\max(X_1,$ $X_2)$ and $X_0$ for Gaussian RVs $X_i$ ($i=0,1,2$), allows one to build a closed-loop algorithm of forward traversal of a delay through a timing graph $G$. The requirement for this is that non-Gaussian PDFs of delays are presented via Gaussian mixtures, sums of RBFs of Gaussian form. The decomposition is equivalent to solving the minimisation problem~\eqref{eq:least_squares}.
    \item The low complexity of the Gaussian comb allows construction of efficient SSTA algorithms. Thus, Azuma\textit{et al.}~\cite{azuma2017ApproximatingMaximumGaussians} also proposed to use GMMs but did not propose a procedure for GMM preparation; Cheng~\textit{et al.}~\cite{cheng2007NonLinearStatisticalStatic} used Fourier Series decomposition for evaluation of PDFs, which required exhaustive numerical computations. This is not the case for the proposed approach in this study.
\end{itemize}

\bibliography{edalib}

\end{document}